\documentclass[onecolumn]{elsart3p}
\usepackage{epsfig}
\usepackage{subfigure}
\usepackage{graphicx}
\usepackage{epstopdf}
\usepackage{comment}
\usepackage{color}

\usepackage{overcite}
\usepackage{graphpap}
\usepackage{indentfirst}

\renewcommand{\thefootnote}{\fnsymbol{footnote}}

\newcommand{\noter}[1]{\textcolor{black}{#1}}

\makeatletter \renewcommand\@biblabel[1]{#1} \makeatother
\let\footnotesize=\small

\usepackage{pslatex}
\begin{document}

\begin{frontmatter}

\title{A Solution Accurate, Efficient and Stable Unsplit Staggered Mesh Scheme for Three Dimensional Magnetohydrodynamics}
\author{Dongwook Lee}
\address{The Flash Center for Computational Science, University of Chicago, 5747 S. Ellis, Chicago, IL 60637}
\ead{dongwook@flash.uchicago.edu}

\begin{abstract}
  \normalsize In this paper, we extend the unsplit staggered mesh scheme (USM) for
  2D magnetohydrodynamics (MHD) [D. Lee, A. Deane, 
  An Unsplit Staggered Mesh Scheme for Multidimensional Magnetohydrodynamics, J. Comput. Phys. 
  228 (2009) 952--975]
  to a full 3D MHD scheme. 
  The scheme is a finite-volume Godunov method consisting of a constrained transport (CT) method 
  and an efficient and accurate single-step, directionally unsplit multidimensional 
  data reconstruction-evolution algorithm, which
  extends Colella's original 2D corner transport upwind (CTU) method [P. Colella, Multidimensional Upwind
  Methods for Hyperbolic Conservation Laws, J. Comput. Phys. 87 (1990) 446--466].
  We present two types of data reconstruction-evolution algorithms for 3D: (1) a reduced CTU scheme and (2) a full CTU scheme.
  \noter{
  The reduced 3D CTU scheme is a variant of a simple 3D extension of Collela's 2D CTU method 
  and is considered as a direct extension from the 2D USM scheme. 
  The full 3D CTU scheme is our primary 3D solver which includes all multidimensional cross-derivative terms
  for stability. The latter method is logically analogous to the 3D unsplit CTU method by Saltzman
  [J. Saltzman, An unsplit 3D upwind method for hyperbolic conservation laws, J. Comput. Phys. 
  115 (1994) 153--168]. 
  The major novelties in our algorithms are twofold. 
  First, we extend the reduced CTU scheme to the full CTU scheme which is able to run with CFL numbers close to unity. 
  Both methods utilize the transverse update technique developed in the 2D USM algorithm to account for 
  transverse fluxes
   {\it{without}} solving intermediate Riemann problems, which in turn gives
  cost-effective 3D methods by reducing the total number of Riemann solves.}
  The proposed algorithms are simple and efficient especially when including
  multidimensional MHD terms 
  that maintain in-plane magnetic field dynamics. 
  Second, we introduce a new CT scheme that makes use
  of proper upwind information in taking averages of electric fields. 
  Our 3D USM schemes can be easily combined with various reconstruction methods
  (e.g., first-order Godunov, second-order MUSCL-Hancock, third-order PPM and
  fifth-order WENO), and a wide choice of \noter{1D based} Riemann solvers
  (e.g., local Lax-Friedrichs, HLLE, HLLC, HLLD, and Roe). 
  The 3D USM-MHD solver is
  available in the University of Chicago Flash Center's official FLASH release.
\end{abstract}


\end{frontmatter}

\large

\section{Introduction}
Many astrophysical applications involve the study of magnetized flows generating shock waves.
Modeling such flows requires numerical solution of the equations of magnetohydrodynamics (MHD)
that couple the magnetic field to the gas hydrodynamics using Maxwell's equations.
A valid computer model needs to capture accurately the nonlinear shock propagation
in the magnetized flows without sacrificing computational efficiency and stability.

Obviously, with suitable assumptions about flow symmetries, a simple approach to
obtain a computationally efficient model is to
solve a reduced system in 1D or 2D instead of 3D. However, a limitation of
such reduced systems is that they cannot be used to understand complicated nonlinear physics
occurring only in  the full 3D situation. Although solving the reduced system can 
illustrate interesting characteristic features (e.g., the inverse energy cascade in 2D turbulence [\citenum{Kraichnan1967}]),
it is essential to use 3D simulations term
in order to understand the full nonlinear nature of MHD phenomena (e.g., the
energy cascade from large scales to small scales in 3D turbulence) .

There are two approaches in modeling multidimensional (i.e., 2D and 3D) algorithms for gas hydrodynamics
and MHD in terms of spatial integration methods: split and unsplit.
The directionally split method has the advantage of extending a 1D algorithm to
higher dimensions, simply by conducting directional sweeps along additional dimensions,
in which each sweep solves 1D sub-system.
Thus, the Courant-Friedrichs-Lewy (CFL) numerical stability constraint of the split schemes in multi-dimensions
is the same as the 1D constraint, which is to say CFL$\le 1.0$.
Despite their simplicity and robustness, however, a number of recent studies 
have revealed numerical problems in the split formulations of multidimensional
MHD and gas hydrodynamics (e.g., loss of expected flow symmetries [\citenum{AlmgrenEtAl2006,LeeAstronum2009}], 
failure to preserve in-plane magnetic field evolution [\citenum{GardinerStone2005, LeeDeane2009}], 
numerical artifacts due to a failure to compute proper strain rates on a grid scale [\citenum{AlmgrenEtAl2010}]).

For MHD the use of an unsplit formulation is more critical than for hydrodynamics. 
This is because the split formulations fail to evolve the normal (in the sweep direction) magnetic field
[\citenum{Crockett2005,GardinerStone2005,GardinerStone2008,StoneGardinerEtAlAthena2008}].
For 2D MHD, Gardiner and Stone [\citenum{GardinerStone2005}]
identified the importance of such multidimensional consideration
in their unsplit MHD scheme based on the corner transport upwind (CTU) \noter{[\citenum{Colella1990}]}
and the constrained transport (CT) [\citenum{EvansHawley1988}] methods.
Later, the authors proposed a 3D unsplit version of an unsplit MHD scheme
in [\citenum{GardinerStone2008}], in which the
extension of the multidimensional MHD terms from their 2D algorithm to 3D
is accomplished at the cost of considerable algorithmic complexity
and a reduced stability limit (CFL $< 0.5$) in their 6-solve CTU+CT algorithm.
It is known in a \noter{CTU-type} 3D unsplit formulation that the full CFL stability limit (i.e., CFL number $\le1.0$) can be recovered
by accounting for intermediate Riemann problems fully, 
requiring 12 Riemann solves per zone per time step [\citenum{Saltzman1994}].
In general, the calculations associated with the Riemann solves are computationally expensive. 
Gardiner and Stone [\citenum{GardinerStone2008}] considered two alternative
options, an expensive 12-Riemann solve yielding the full CFL limit and
a reduced 6-Riemann solve 
with a more constraining CFL condition  (CFL number $<0.5$). They found that
the two approaches are similar in terms of computational cost and there is
no significant difference in performance between them. 
The 6-solve scheme is chosen to be their primary 3D integrator because of its relatively low complexity
in incorporating the multidimensional MHD terms.

\noter{The CTU formulation has an advantage in its compact design of one-step temporal update which is well-suited 
for multidimensional problems.  
However, it is limited to second-order.
There has been much progress in other types of temporal update strategies that are higher than second-order accurate, taking a different
path from CTU.
Early attempts have utilized a Runge-Kutta (RK) based temporal update formulation coupled with spatially high-order reconstruction schemes in the finite-difference framework [\citenum{Harten1987,ShuOsher1988,ShuOsher1989,BarthFrederickson1990,SureshHuynh1997,Liu1994,JiangShu1996,BalsaraShu2000}]. 
Such RK-based high-order schemes have been also developed in the finite-volume framework  [\citenum{BalsaraEtAl2007,HuShu1999,ZhangShu2003,Dumbser2007a,Dumbser2007b}]
which has superior properties to that of finite-difference for
resolving compressible flows on both uniform and AMR grids.
The high-order RK temporal update strategies rely on multi-stage updates which add to the computational cost.
Therefore it is desirable to retain a CTU-like one-step formulation, while retaining higher than second order accuracy.
Recent work has been found to provide such efficiency using a new formulation so-called the Arbitrary Derivative Riemann Problem (ADER), see [\citenum{TitarevToro2002,TitarevToro2005,ToroTitarev2006,Dumbser2008a,Dumbser2008b,Balsara2009,Balsara2012b}].}
\noter{
For solving multidimensional conservation laws, there has been another line of progress that tries to build genuinely multidimensional
Riemann solvers for hydrodynamics [\citenum{Abgrall1994,Fey1998a,Fey1998b,Gilquin1993,Wendroff1999,Brio2001}].
Recently, a family of two-dimensional HLL-type Riemann solvers, HLLE  [\citenum{Balsara2010}] and HLLC  [\citenum{Balsara2012}], have been introduced and generalized by Balsara for both hydrodynamics and MHD. As shown in his work the multidimensional Riemann solvers are genuinely derived for 2D. A major improvement in MHD flows is that
they inherently provide proper amount of numerical dissipation
 that is necessary to propagate magnetic fields in a stable manner. Alternatively, 1D Riemann solver formulations such as [\citenum{LondrilloDelZanna2004,GardinerStone2005}] need to add extra dissipation for a stable upwinding.
The use of multidimensional Riemann solvers is also shown to capture isotropic wave propagations better than the usual 1D approach. Furthermore, both types of solvers have been extended to 3D using a one-step predictor-corrector formulation.}

\noter{
The above mentioned strategies using high-order schemes and genuinely multidimensional Riemann solvers,
provide improved solution accuracy and stability over CTU-CT formulations. 
In this paper, however, we are primarily interested in constructing a scheme that can be built on the
1D Riemann solver framework in line with a CTU-type method. The latter is 
 (arguably) most widely used in many Godunov-type modern codes 
 [\citenum{Colella1990,Saltzman1994,Li2003,LondrilloDelZanna2004,
 GardinerStone2005,GardinerStone2008,LeeDeane2009,MignoneTzeferacos2010,MiniatiMartin2011}]. 
This design also benefits us in extending 
our 2D USM-MHD algorithm [\citenum{LeeDeane2009}]  to 3D without any modifications of the Riemann solvers.}
This paper describes an approach that provides 
(i) an algorithmic extension from 2D to 3D of the USM scheme of 
Lee and Deane [\citenum{LeeDeane2009}], 
and (ii) the full CFL stability bound in 3D {\it{without}} the expense of 12 Riemann solves per cell per time step,
and (iii) a new upwind biased electric fields construction scheme for CT. 
We show that the present USM scheme achieves a
numerically efficient and consistent MHD algorithm in 3D
without introducing a greater amount of additional complexity, while maintaining the full CFL stability
range.

The paper is organized as follows: Section \ref{section2}
describes our new 3D unsplit, single-step 
\noter{data reconstruction-evolution USM algorithm which consists of two stages, i.e., normal predictor and
transverse corrector.}
Section \ref{section2} is subdivided
into several subsections.  We begin in Section
\ref{sec:MHDsystem} our discussion of the 3D USM scheme by considering the
governing equations of MHD and their linearized form.
The second-order MUSCL-Hancock approach for calculating the normal predictor
is described in Section \ref{sec:normal-predictor-dataReconst}.
We introduce in 
Section \ref{sec:TransverseCorrector}
our two 3D CTU schemes to compute the transverse correctors,
which are efficient and essential for obtaining the full CFL stability range. In
the subsections therein, we construct Riemann states at cell
interfaces, focusing on our new transverse correctors that do not require
the solution of any Riemann problem. The Riemann state calculations
are completed  by evolving the normal magnetic fields by a half time step, about which
we describe in Section \ref{sec:CThalfTimeStepUpdate}.
The final update of the cell-centered
conservative variables is shown in Section \ref{sec:unkUpdate}, followed by
a new 3D upwind-biased CT update of magnetic fields
in Section \ref{sec:MEC}.
We summarize our step-by-step, point-to-point 3D CTU schemes in Section \ref{sec:summary}.
In Section \ref{sec:results} we present numerical results of various test problems that demonstrate the
qualitative and quantitative performance of our schemes. 
We conclude the paper with a discussion in Section \ref{sec:conclusion}.

\section{The three-dimensional USM scheme for MHD}
\label{section2}

\subsection{MHD Equations}
\label{sec:MHDsystem}
We consider solving the equations of MHD in conservation form
\begin{eqnarray}
&&\frac{\partial \rho}{\partial t}+\nabla \cdot \left(\rho\mathbf u\right)=0,\label{sec2:ContinuityEqn} \\
&&\frac{\partial \rho\mathbf u}{\partial t}+
\nabla \cdot \left(\rho\mathbf u \mathbf u - \mathbf B \mathbf B \right)
+\nabla p_{tot} =0, \label{sec2:MomentumEqn} \\
&&\frac{\partial \mathbf B}{\partial t}+
\nabla \cdot \left(\mathbf u \mathbf B -\mathbf B \mathbf u \right)=0, \label{sec2:InductionEqn}\\
&&\frac{\partial E}{\partial t}+
\nabla \cdot \left(\mathbf u E +\mathbf u p_{tot}-\mathbf B\mathbf B\cdot \mathbf u \right)=0.
\label{sec2:EnergyEqn}
\end{eqnarray}
The conservative variables include the
plasma mass density $\rho$, momenta $\rho\mathbf u$, magnetic fields
$\mathbf B$, and total energy density $E$. 
\noter{
The rest are the thermal pressure $p=(\gamma-1)(E-\frac{1}{2}\rho U^2-B_p)$, the magnetic
pressure $B_p=(B_x^2+B_y^2+B_z^2)/2$, and the sum of the two is the total pressure
$p_{tot}=p+B_p$. The ratio of specific heats is
denoted with  $\gamma$  as usual. 
The solenoidal constraint $\nabla \cdot\mathbf B=0$ is implied
in the induction equation.
}

We write the above equations in a matrix form in 3D
\begin{equation}
\frac{\partial \mathbf U}{\partial t}+ \frac{\partial\mathbf F}{\partial x} + \frac{\partial\mathbf G}{\partial y} + \frac{\partial\mathbf H}{\partial z}=0,
\label{sec2:GovEqn2}
\end{equation}
where $\mathbf{U}$ contains the eight MHD conservative variables, and
$\mathbf{F}$, $\mathbf{G}$, and $\mathbf{H}$ 
represent the corresponding conservative fluxes
in $x,y$ and $z$ directions.     
It is often convenient
to cast the conservative form of Equation (\ref{sec2:GovEqn2}) into a
quasi-linearized representation in terms of primitive variables,
$\mathbf V =\bigr(\rho, u,  v, w, B_x, B_y, B_z, p\bigl)^T$,
in order to discretize the coupled system of MHD equations
(\ref{sec2:ContinuityEqn})-(\ref{sec2:EnergyEqn}), 
\begin{eqnarray}
\frac{\partial \mathbf V}{\partial t}+
{\mathbf A}_x\frac{\partial \mathbf V}{\partial x} + 
{\mathbf A}_y\frac{\partial \mathbf V}{\partial y} +
{\mathbf A}_z\frac{\partial \mathbf V}{\partial z}=0.
\label{sec2:PrimGovExact}
\end{eqnarray}
The coefficient matrices ${\mathbf A}_x$, ${\mathbf A}_y$, and ${\mathbf A}_z$ are given by
\begin{eqnarray}
\label{sec2:Ax}
&&{\mathbf A}_x
=\pmatrix{u &\rho    &  0      &  0      & 0               &  0              & 0               & 0            \cr 
          0 & u      &  0      &  0      &-\frac{B_x}{\rho}&\frac{B_y}{\rho} & \frac{B_z}{\rho}&\frac{1}{\rho}\cr
          0 & 0      &  u      &  0      &-\frac{B_y}{\rho}&-\frac{B_x}{\rho}& 0               & 0            \cr
          0 & 0      &  0      &  u      &-\frac{B_z}{\rho}&  0              &-\frac{B_x}{\rho}& 0            \cr
	  0 & 0      &  0      &  0      & 0               &  0              & 0               & 0            \cr
          0 & B_y    &-B_x     &  0      &-v               &  u              & 0               & 0            \cr
          0 & B_z    &  0      &-B_x     &-w               &  0              & u               & 0            \cr
          0 &\gamma p&  0      &  0      &-k{\mathbf u \cdot \mathbf B}
                                                           &  0              & 0               & u            \cr },\\ 
\label{sec2:Ay}
&&{\mathbf A}_y
=\pmatrix{v & 0      &\rho     &  0      & 0               &  0              & 0               & 0            \cr 
          0 & v      &  0      &  0      &-\frac{B_y}{\rho}&-\frac{B_x}{\rho}& 0               & 0            \cr
          0 & 0      &  v      &  0      & \frac{B_x}{\rho}&-\frac{B_y}{\rho}&\frac{B_z}{\rho} &\frac{1}{\rho}\cr
          0 & 0      &  0      &  v      & 0               &-\frac{B_z}{\rho}&-\frac{B_y}{\rho}& 0            \cr
          0 &-B_y    & B_x     &  0      & v               &-u               & 0               & 0            \cr
	  0 & 0      &  0      &  0      & 0               & 0               & 0               & 0            \cr
	  0 & 0      & B_z     &-B_y     & 0               &-w               & v               & 0            \cr
          0 & 0      &\gamma p &  0      & 0               &-k{\mathbf u \cdot \mathbf B}
                                                                             & 0               & v            \cr },\\ 
\label{sec2:Az}
&&{\mathbf A}_z
=\pmatrix{w & 0      &  0      &\rho     & 0               &  0              & 0               & 0            \cr 
          0 & w      &  0      &  0      &-\frac{B_z}{\rho}&  0              &-\frac{B_x}{\rho}& 0            \cr
          0 & 0      &  w      &  0      & 0               &-\frac{B_z}{\rho}&-\frac{B_y}{\rho}& 0            \cr
          0 & 0      &  0      &  w      & \frac{B_x}{\rho}& \frac{B_y}{\rho}&-\frac{B_z}{\rho}&\frac{1}{\rho}\cr
          0 &-B_z    &  0      & B_x     & w               & 0               &-u               & 0            \cr
 	  0 & 0      &-B_z     & B_y     & 0               & w               &-v               & 0            \cr 
	  0 & 0      &  0      &  0      & 0               & 0               & 0               & 0            \cr
          0 & 0      &  0      &\gamma p & 0               & 0               &-k{\mathbf u \cdot \mathbf B}
                                                                                               & w            \cr }, 
\end{eqnarray}
with $k=1-\gamma$. 
 
For exposition purposes in this paper, we illustrate our calculations using a spatially second-order
MUSCL-Hancock (MH) piecewise-linear method (PLM) for the normal predictor.
Other normal predictor algorithms
(e.g., piecewise parabolic method (PPM [\citenum{ColellaWoodward1984}]), 
essentially non-oscillatory (ENO [\citenum{HartenENO1987}]),
weighted essentially non-oscillatory (WENO [\citenum{JiangShu1996}]), etc.)
can be adopted as well to give
different degrees of solution accuracy in our algorithm. In fact, we have
implemented various reconstruction schemes of MH, PPM and 5th order WENO in FLASH, and they are available in
the official FLASH distribution
[\citenum{LeeDeane2009,LeeAstronum2009,Flash,DubeyFLASH2009}].

\begin{figure}[htpb]
\begin{picture}(200,220)(20,0)
\put(190,0){\line(0,1){210}}
\put(325,0){\line(0,1){210}}
\put(150,35){\line(1,0){210}}
\put(150,175){\line(1,0){210}}
\put(252,102){$\ast$}\put(256,100){$(i,j)$}
\put(250,20) {$\mathbf V^{n+1/2}_{i,j-1,N}$}
\put(250,45) {$\mathbf V^{n+1/2}_{i,j,S}$}
\put(250,155){$\mathbf V^{n+1/2}_{i,j,N}$}
\put(250,185){$\mathbf V^{n+1/2}_{i,j+1,S}$}
\put(148,100){$\mathbf V^{n+1/2}_{i-1,j,E}$}
\put(195,100){$\mathbf V^{n+1/2}_{i,j,W}$}
\put(288,100){$\mathbf V^{n+1/2}_{i,j,E}$}
\put(330,100){$\mathbf V^{n+1/2}_{i+1,j,W}$}
\end{picture}

\centering
\caption{The boundary extrapolated values on a 2D cell geometry.
Our subscriptions $N,S,E,W,T,B$ represent respectively
north, south, east, west, top and bottom that are based on a reference
point at the local cell center node $(i,j,k)$.
}
\label{sec2:BoundaryExtrapolation}
\end{figure}
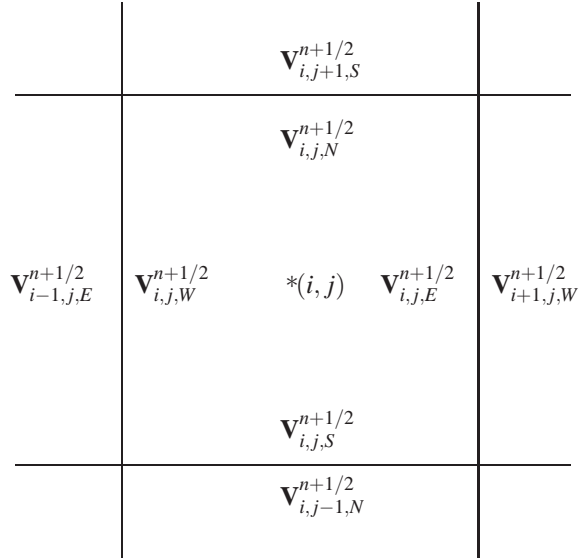

This brings us to write the system (\ref{sec2:PrimGovExact}) to obtain second-order accurate discretizations
at cell faces,
\begin{eqnarray}
&&\mathbf{{V}}_{i,j,k,E,W}^{n+1/2}=
\mathbf{{V}}_{i,j,k}^n+
\frac{1}{2}\bigl[\pm\mathbf{I}-\frac{\Delta t}{\Delta x}\mathbf{{A}}_x\bigr]\Delta_{x}^{tvd} \mathbf{V}^{n}_{i,j,k}
-\frac{\Delta t}{2\Delta y}\mathbf{{A}}_y\Delta_{y}^{up} \mathbf{V}^{n}_{i,j,k}
-\frac{\Delta t}{2\Delta z}\mathbf{{A}}_z\Delta_{z}^{up} \mathbf{V}^{n}_{i,j,k},
\label{sec2:Evolx_2}\\
&&\mathbf{{V}}_{i,j,k,N,S}^{n+1/2}=
\mathbf{{V}}_{i,j,k}^n
-\frac{\Delta t}{2\Delta x}\mathbf{{A}}_x\Delta_{x}^{up} \mathbf{V}^{n}_{i,j,k}
+\frac{1}{2}\bigl[\pm\mathbf{I}-\frac{\Delta t}{\Delta y}\mathbf{{A}}_y\bigr]\Delta_{y}^{tvd} \mathbf{V}^{n}_{i,j,k}
-\frac{\Delta t}{2\Delta z}\mathbf{{A}}_z\Delta_{z}^{up} \mathbf{V}^{n}_{i,j,k},
\label{sec2:Evoly_2}\\
&&\mathbf{{V}}_{i,j,k,T,B}^{n+1/2}=
\mathbf{{V}}_{i,j,k}^n
-\frac{\Delta t}{2\Delta x}\mathbf{{A}}_x\Delta_{x}^{up} \mathbf{V}^{n}_{i,j,k}
-\frac{\Delta t}{2\Delta y}\mathbf{{A}}_y\Delta_{y}^{up} \mathbf{V}^{n}_{i,j,k},\
+\frac{1}{2}\bigl[\pm\mathbf{I}-\frac{\Delta t}{\Delta z}\mathbf{{A}}_z\bigr]\Delta_{z}^{tvd} \mathbf{V}^{n}_{i,j,k},
\label{sec2:Evolz_2}
\end{eqnarray}
where the plus and minus signs correspond to directions of $N,E,S,W,T$ and
$B$ respectively in a natural way, see Figure \ref{sec2:BoundaryExtrapolation}.
Each $\mathbf{{A}}_d$
matrix represents the
coefficient matrix in the $d$-direction evaluated at $\mathbf{{V}}_{i,j,k}^n$.
\noter{
The undivided difference operators in each $d$-direction are denoted as ${\Delta}_{d}^{tvd}$ and ${\Delta}_{d}^{up}$, 
and they are
suitably chosen slope vectors of $\mathbf{{V}}_{i,j,k}^n$ in each cell $(i,j,k)$ using 
TVD and upwind slope limiters, respectively.
}

\subsection{Normal Predictor}
\label{sec:normal-predictor-dataReconst}
The first stage
is to calculate the normal predictor states,
including all the required multidimensional MHD terms (the MHD terms hearafter) 
[\citenum{LeeDeane2009}] satisfying the solenoidal constraint $\nabla \cdot
\mathbf{B} = 0$. 
We begin our discussion with the evolution of the normal field, $B_N$,
which is treated separately from the other primitive variables. 
For instance, when $N=x$, we can define
\begin{equation}
{\mathbf {\bar{V}}}_x=\left[\begin{array}{c}
                        {\mathbf {\hat {V}}}_x\\
          	        B_x
			\end{array}
                \right]
\mbox{ and }
{\mathbf {\bar{A}}}_x=\left[\begin{array}{cc}
			\mathbf {\hat {A}}_x & \mathbf {A}_{B_x} \\
 			\mathbf {0}        & 0
			\end{array}
		  \right].
\label{sec2:ReducedSystem_x}
\end{equation}
Here ${\mathbf {\hat{V}}}_x$ is a $7\times1$ vector excluding $B_x$,
$\mathbf {\hat {A}}_x$ is a $7\times7$ matrix omitting both the fifth row
and column in the original matrix $\mathbf {A}_x$ in Equation (\ref{sec2:Ax}), and
$\mathbf {A}_{B_x}$ is a $7\times1$ vector,
\begin{equation}
\mathbf {A}_{B_x}=\left[0, -\frac{B_x}{\rho}, -\frac{B_y}{\rho}, -\frac{B_z}{\rho}, -v,-w, -k{\mathbf u \cdot \mathbf B} \right]^T.
\label{sec2:Abx}
\end{equation}
Note that the hat
(\verb|^|) notation denotes the reduced system (i.e., the
one corresponding to the usual 1D MHD equations) and the bar (\verb|-|)
notation indicates the re-assembled full system.
Similarly for the other directions, we have
\begin{equation}
{\mathbf {\bar{V}}}_y=\left[\begin{array}{c}
                        {\mathbf {\hat {V}}}_y\\
          	        B_y
			\end{array}
                \right], \;\;\;
{\mathbf {\bar{A}}}_y=\left[\begin{array}{cc}
			\mathbf {\hat {A}}_y & \mathbf {A}_{B_y} \\
 			\mathbf {0}        & 0
			\end{array}
		  \right], \;\;\;
\mathbf {A}_{B_y}=\left[0, -\frac{B_x}{\rho}, -\frac{B_y}{\rho}, -\frac{B_z}{\rho}, -u,-w, -k{\mathbf u \cdot \mathbf B} \right]^T,
\label{sec2:ReducedSystem_y}
\end{equation}

\begin{equation}
{\mathbf {\bar{V}}}_z=\left[\begin{array}{c}
                        {\mathbf {\hat {V}}}_z\\
          	        B_z
			\end{array}
                \right], \;\;\;
{\mathbf {\bar{A}}}_z=\left[\begin{array}{cc}
			\mathbf {\hat {A}}_z & \mathbf {A}_{B_z} \\
 			\mathbf {0}        & 0
			\end{array}
		  \right], \;\;\;
\mathbf {A}_{B_z}=\left[0, -\frac{B_x}{\rho}, -\frac{B_y}{\rho}, -\frac{B_z}{\rho}, -u,-v, -k{\mathbf u \cdot \mathbf B} \right]^T.
\label{sec2:ReducedSystem_z}
\end{equation}
The term $\mathbf {A}_{B_N}$ for each $N$ will be our representation of the corresponding
MHD term in this paper. 

The first step of MH
extrapolates $\mathbf V^{n}_{i,j,k}$ to construct 
the six multidimensional Riemann states 
$\mathbf V^{n+1/2}_{i,j,k,N,S,E,W,T,B}$ 
at cell interfaces
to achieve second-order accuracy by using a total variation diminishing (TVD) 
slope limiter\footnote{For instance, limiters such as minmod, van Leer's, 
monotonized central (MC), or 
a hybrid combination of them on different wave structures [\citenum{Balsara2003}]
can be used.}. Although the slope limiter can be applied to either
primitive or characteristic variables, we prefer
the latter
since it is less prone to generating spurious oscillations
as noted in the literature
[\citenum{Toro2009, vanLeer2006}].
We do not apply any limiting to $B_N$,
allowing the 
continuity of the normal field at cell faces 
(e.g., see discussion in [\citenum{LeeDeane2009}]).
To simplify our discussion, we focus on
the $x$-direction in Equation (\ref{sec2:Evolx_2}). The others in Equations
(\ref{sec2:Evoly_2})-(\ref{sec2:Evolz_2}) can be computed in the similar
way. 
We consider the first two terms in (\ref{sec2:Evolx_2})
that are related to the normal predictor
\begin{equation}
\left[\begin{array}{c} {\mathbf {\hat {V}}}_x\\B_x\end{array}\right]_{i,j,k,E,W}^{n+1/2,\parallel}=
\left[\begin{array}{c} {\mathbf {\hat {V}}}_x\\B_x\end{array}\right]_{i,j,k}^{n}+\frac{1}{2}
\left(\pm
\left[\begin{array}{cc}\mathbf {\hat {I}}   & \mathbf {0}       \\ \mathbf {0}& 1 \end{array} \right]
-\frac{\Delta t}{\Delta x}
\left[\begin{array}{cc}\mathbf {\hat {A}}_x & \mathbf {A}_{B_x} \\ \mathbf {0}& 0 \end{array} \right]_{i,j,k}^n
\right)
{\Delta}_{x}^{tvd}\bar\mathbf{V}^{n}_{i,j,k},
\label{sec2:NormalPredictorFull}
\end{equation}
where 
${\Delta}_{x}^{tvd}\bar\mathbf{V}^{n}_{i,j,k}=\Bigl({\Delta}_{x}^{tvd}\hat\mathbf{V}^{n}_{i,j,k},
\Delta B_{x,i}^n\Bigr)^T$ and $\Delta B_{x,i}^n=b_{x,i+1/2,j,k}^n-b_{x,i-1/2,j,k}^n$.
The notations $B_d$ and
$b_d$ denote cell-centered and cell face-centered magnetic fields
respectively, with $d=x,y,z$. In CT, $\Delta
B_{x,i}^n$ is constructed such that the numerical divergence is zero using
the cell face-centered magnetic fields. In other words, $\Delta B_{x,i}^n$,
$\Delta B_{y,j}^n$ and $\Delta B_{z,k}^n$ are chosen such that
\begin{equation}
\frac{\Delta B_{x,i}^n}{\Delta x} + \frac{\Delta B_{y,j}^n}{\Delta y} + \frac{\Delta B_{z,k}^n}{\Delta z} =0,
\label{sec2:DivFreeConstraint}
\end{equation}
where we analogously define $\Delta B_{y,j}^n$ and $\Delta B_{z,k}^n$. 
Solving a system in relation (\ref{sec2:NormalPredictorFull}) is equivalent to considering two sub-systems
\begin{equation}
\cases{
\mathbf {\hat {V}}_{x,i,j,k,E,W}^{n+1/2,\parallel}=
\mathbf {\hat {V}}_{x,i,j,k}^n +\frac{1}{2}\left(
\pm \mathbf {\hat {I}} -\frac{\Delta t}{\Delta x}\mathbf {\hat {A}}_x \right)_{i,j,k}^n{\Delta}_{x}^{tvd}\hat\mathbf{V}^{n}_{i,j,k}
                       -\frac{\Delta t}{2\Delta x}(\mathbf {A}_{B_x})_{i,j,k}^n \Delta B^n_{x,i},\cr
\left(B_x\right)_{i,j,k,E,W}^{n+1/2,\parallel} =
B_{x,i,j,k}^{n} \pm \frac{1}{2}\Delta B_{x,i}^n. \cr}
\label{sec2:NormalPredictorReduced}
\end{equation}
The second relation in (\ref{sec2:NormalPredictorReduced}) is nothing but
\begin{equation}
\left(B_x\right)_{i,j,k,E,W}^{n+1/2,\parallel}=B_{x,i,j,k}^{n} \pm \frac{1}{2}\Delta B_{x,i}^n=b_{x,i\pm1/2,j,k}^n,
\label{sec2:BdryExtraBx}
\end{equation}
because we use a simple arithmetic averaging to obtain the cell-centered
magnetic field using the divergence-free fields at cell interface centers,
\begin{equation}
\label{sec2:BxCenterReconstruction}
B_{x,i,j,k}^n=\frac{1}{2}\Bigl(b_{x,i+1/2,j,k}^n+b_{x,i-1/2,j,k}^n\Bigr).
\end{equation}

Applying the characteristic tracing method in $x$-normal direction in (\ref{sec2:NormalPredictorReduced}) yields

\begin{equation}
\label{WestNormalUpdate}
{\mathbf {\hat {V}}}_{x,i,j,k,W}^{n+1/2,\parallel}=
\hat {\mathbf V}_{x,i,j.k}^n+
\frac{1}{2}\sum_{m;\lambda^m_{x,i,j,k}<0}\Bigl(-1-\frac{\Delta t}{\Delta x}\lambda^m_{x,i,j,k}\Bigr)
\mathbf{r}^m_{x,i,j,k}
{\Delta}^{tvd}_x\hat\alpha^{n}_{i,j,k}
-\frac{\Delta t}{2\Delta x}(\mathbf {A}_{B_x})_{i,j,k}^n \Delta B^n_{x,i},
\end{equation}
\begin{equation}
\label{EastNormalUpdate}
\mathbf {\hat {V}}_{x,i,j,k,E}^{n+1/2,\parallel}=
\hat {\mathbf V}_{x,i,j,k}^n+
\frac{1}{2}\sum_{m;\lambda^m_{x,i,j,k}>0}\Bigl(1-\frac{\Delta t}{\Delta x}\lambda^m_{x,i,j,k}\Bigr)
\mathbf{r}^m_{x,i,j,k}
{\Delta}^{tvd}_x\hat\alpha^{n}_{i,j,k}
-\frac{\Delta t}{2\Delta x}(\mathbf {A}_{B_x})_{i,j,k}^n \Delta B^n_{x,i}.
\end{equation}
A suitable TVD slope limiter along the $x$-normal direction
is used in the undivided slope operator on each characteristic variable $\hat\alpha$ 
\begin{equation}
\label{sec2:TvdLimiterCharX}
{\Delta}^{tvd}_x\hat\alpha^{n}_{i,j,k}
={\verb|TVD_Limiter|}
\Bigl[\mathbf{l}^m_{x,i,j,k}\cdot (\mathbf{\hat{V}}^n_{x,i+1,j,k}-\mathbf{\hat{V}}^n_{x,i,j,k}),
         \mathbf{l}^m_{x,i,j,k}\cdot (\mathbf{\hat{V}}^n_{x,i,j,k}-\mathbf{\hat{V}}^n_{x,i-1,j,k})
         \Bigr].
\end{equation}
Here $\lambda^m_{x,i,j,k}, \mathbf{r}^m_{x,i,j,k}, \mathbf{l}^m_{x,i,j,k}$
represent respectively the eigenvalue, right and left eigenvectors of
$\mathbf {\hat {A}}_x$, calculated at the corresponding cell center
$(i,j,k)$ in the $x$-direction at time step $n$.
This completes the first part of our description on a single-step, data
reconstruction-evolution algorithm in the $x$-normal direction. 

\subsection{Transverse Corrector in USM}
\label{sec:TransverseCorrector}
\subsubsection{Review of Computing Transverse Flux Gradients using Characteristic Tracing}
\label{sec:CharacteristicTracingForTransverse}
The transverse corrector adds the gradients of transverse
fluxes to the normal predictors. This transverse
corrector step plays a crucial role for stability in CTU.
Generally speaking, the degree of
accuracy is affected by the normal predictor, whereas numerical stability is
strongly determined by the transverse corrector [\citenum{Bell1988}].

In [\citenum{LeeDeane2009}], Lee and Deane noted that the transverse
flux gradients, responsible for the cross-derivative terms in CTU,
which assure stability for flows advecting along diagonal corner
directions, can be replaced by a simpler approach that is based on
characteristic tracing alone. 
This removes the need to solve the
intermediate Riemann problems. As a result, this approach requires only two Riemann solutions in 2D
(not counting the extra two Riemann solves to update the divergence-free
magnetic fields by CT), while preserving the full stability of the CTU
scheme.
We review a pointwise description of the transverse corrector in USM for a moment. 
Consider the $y$-transverse flux gradient (i.e., the third term in (\ref{sec2:Evolx_2})) 
which supplies the corrector term to the $x$-normal predictor states. 
For any left ($\mathbf {\hat{V}}_{i,j-,k}$) and right ($\mathbf {\hat{V}}_{i,j+,k}$) states at cell $(i,j,k)$ along $y$-direction,
the jump conditions across the
individual $m$-th wave gives
\begin{equation}
\label{Eqn-section2:JumpRelationship}
\mathbf {\hat {A}}_{y,i,j,k}\mathbf {\hat{V}}_{y,i,j-,k} + \sum_{m=1}^{m_0-1} \lambda^m_{y,i,j,k} {\mathbf {r}}^m_{y,i,j,k} {\Delta}^{up}_y\hat\alpha^{n}_{i,j,k}
=\mathbf {\hat {A}}_{y,i,j,k}\mathbf {\hat{V}}_{y,i,j+,k} - \sum_{m=m_0}^{7} \lambda^m_{y,i,j,k} {\mathbf {r}}^m_{y,i,j,k} {\Delta}^{up}_y\hat\alpha^{n}_{i,j,k}.
\end{equation}
\noter{Now recall that the property of conservation [\citenum{Toro2009,LeVeque1992}] across discontinuities of 
the Roe matrix $\mathbf{A}$. 
It states that the Roe matrix ensures conservation across 
a discontinuity between the left ($\mathbf{V}_l$) and right ($\mathbf{V}_r$)  states, given by
$\mathbf{Flux}(\mathbf {{V}}_r)-\mathbf{Flux}(\mathbf {{V}}_l)=
\mathbf {A}(\mathbf  {{V}}_r - \mathbf {{V}}_l).$
Applying this relation to $\mathbf {\hat {A}}_{y,i,j,k}$, $\mathbf {\hat{V}}_{y,i,j-,k}$ and $\mathbf {\hat{V}}_{y,i,j+,k}$, and
from (\ref{Eqn-section2:JumpRelationship}), we obtain 
\begin{equation}
\label{Eqn-section2:TransFluxGradients}
{\mathbf{G}}_{i,j+1/2} -  {\mathbf{G}}_{i,j-1/2}= 
 \mathbf {\hat {A}}_{y,i,j,k}(\mathbf {\hat{V}}_{y,i,j+,k}-\mathbf {\hat{V}}_{y,i,j-,k})=
 \sum_{m=1}^{7} \lambda^m_{y,i,j,k} {\mathbf {r}}^m_{y,i,j,k}  {\Delta}^{up}_y\hat\alpha^{n}_{i,j,k}.
\end{equation}
The upwind slope limiter ${\Delta}_{y}^{up}$ is applied to each characteristic variable $\hat\alpha_{i,j,k}^n$ as
\begin{eqnarray}
\label{Eqn-section2:upwindSlopeLimiter}
{\Delta}^{up}_y\hat\alpha^{n}_{i,j,k}
\cases{\mathbf{l}^m_{y,i,j,k}\cdot (\mathbf{\hat{V}}^n_{i,j+1,k}-\mathbf{\hat{V}}^n_{i,j,k}) \;\;\;\; \mbox{if} \;\;\; \lambda^m_{y,i,j,k} < 0\cr
             \mathbf{l}^m_{y,i,j,k}\cdot (\mathbf{\hat{V}}^n_{i,j,k}-\mathbf{\hat{V}}^n_{i,j-1,k})  \;\;\;\; \mbox{if} \;\;\; \lambda^m_{y,i,j,k} >0}.
\end{eqnarray}}

In Equation (\ref{Eqn-section2:TransFluxGradients}) we see that the sum over all
wave contributions gives an effective upwinding of transverse flux
gradients in $y$-direction. The advantage in this approach is that there is no need to solve
the intermediate Riemann problems to add the transverse flux gradient
correction terms to the spatially reconstructed, temporally evolved, normal
predictor states in order to gain the upwind stability in the CTU formulation. 
Because we rely on using the eigensystem in Equation
(\ref{Eqn-section2:TransFluxGradients}), one might suspect that this
characteristic tracing approach could be as expensive as
directly solving the associated Riemann problems at each
interface, followed by taking the gradient of the computed transverse
fluxes. However, this is not the case because we reuse the $y$- (or $x$-)
directional eigensystems that were already calculated in the
normal predictor step in the $y$- (or $x$-) direction. Thus there
is no need to compute any additional eigenstructure for each transverse
direction, which makes our scheme much more computationally efficient than
the standard CTU method. A Fortran-like pseudo code illustrating the algorithm is as follows:

\begin{verbatim}

do j=jmin,jmax
     do i=imin,imax

     ! Compute normal predictor in x-direction, and
     ! store x-directional normal predictor states & eigensystems in arrays
     call dataReconstructNormalDirection(x_dir, x_normalPredictStates, sigmaSum_x)
     
      ! Compute normal predictor in y-direction, and
      ! store y-directional normal predictor states & eigensystems in arrays
     call dataReconstructNormalDirection(y_dir, y_normalPredictStates, sigmaSum_y)

     !  Transverse Correction to the x-normal predictor 
     x_normalPredictStates = x_normalPredictStates - 0.5*dt/dy*sigmaSum_y
     
     !  Transverse Correction to the y-normal predictor 
     y_normalPredictStates = y_normalPredictStates - 0.5*dt/dx*sigmaSum_x
     
     end do
end do

\end{verbatim}
In the above, the terms {\verb|sigmaSum_x|} and {\verb|sigmaSum_y|}
represent the summation of all wave contributions in the $x$- and $y$-directions, 
respectively, given in Equation (\ref{Eqn-section2:TransFluxGradients}).
The rest of the terms are self-explanatory. 

Our approach to approximate the transverse flux gradients, solely using the
characteristic tracing, greatly simplifies the overall unsplit CTU algorithm
by reducing the number of required Riemann solves. In gas hydrodynamics, the
proposed algorithm requires a total of three Riemann solves to update the
solution from $n$ to $n+1$ without compromising
solution stability and accuracy. It will be shown for
MHD in Section \ref{sec:CThalfTimeStepUpdate} that three additional Riemann
problems (yielding a total of six) are required to update the divergence-free, cell
face-centered magnetic fields using the CT method. Another advantage in our
approach, especially for MHD, is the relatively simple handling of
multidimensional MHD terms. This is because our method of adding
transverse flux gradients provides a single-step, directionally unsplit
data reconstruction-evolution algorithm to calculate Riemann states at cell
interfaces. It is therefore much simpler to enforce the balance between flux
gradients in all three directions associated with the MHD terms. As
noted in [\citenum{GardinerStone2008}], complications arise in the
standard full 12-solve CTU scheme, in which the MHD term
balance seems to be hard to achieve
in a series of partial transverse flux gradient
updates based on dimensional splitting.

\subsubsection{Reduced 3D CTU Scheme in USM: Interface State Update from $n$ to $n+1/2$ Time Step}
\label{sec:reduced3dCTU-usm}
Our first simple algorithm using the transverse corrector technique in the previous section 
is analogous to the
6-solve CTU in [\citenum{GardinerStone2008}]. 
This approach can be viewed as a straightforward 3D extension of the
2D CTU scheme [\citenum{Colella1990}], omitting all the third-order cross-derivative terms such as
${\partial^3}/{\partial_x \partial_y \partial_z}$, while including the
second-order cross-derivative terms that are provided in the 2D CTU method.
The resulting Riemann state calculations account for flow information along
the edge directions, but do not fully account for flow information along
the diagonal corner directions, yielding a formal stability
limit of CFL number less than 0.5 \footnote{One can easily prove this stability
bound numerically for a 3D scalar advection equation using a standard von
Neuman Fourier analysis, assuming a single Fourier mode solution
$q^n_{I,J,K}=e^{i(\xi I \Delta x + \eta J \Delta y + \zeta K \Delta z)}$
where $i=\sqrt{-1}$; $I,J,K$ as the grid indices; and $\xi, \eta, \zeta$
the wave numbers in $x,y,z$-directions respectively.}.
This simple approach, referred to as the reduced 3D CTU scheme, can be directly extended from the 2D
CTU [\citenum{LeeDeane2009}] by adding the third additional
transverse flux correction in $z$. That is, the $x$-normal predictors in
Equations (\ref{WestNormalUpdate})-(\ref{EastNormalUpdate}) are
further corrected by including the transverse flux contributions from
$y,z$-directions using the characteristic tracing approach described in the previous section, see also
[\citenum{LeeDeane2009}]. 
For instance, in Equation (\ref{sec2:Evolx_2}) the transverse corrector step can be
updated, first by accounting for the $y$-transverse flux correction,
\begin{equation}
\label{eqn:y-transverse}
\mathbf{V}_{i,j,k,E,W}^{n+1/2,y}=
\mathbf{V}_{i,j,k,E,W}^{n+1/2,\parallel}
-\frac{\Delta t}{2\Delta y}\mathbf{A}_y(\mathbf{V}_{i,j,k}^n)\Delta_{y}^{up}\mathbf{V}_{i,j,k}^{n},
\end{equation}
followed by the $z$-transverse flux correction,
\begin{equation}
\label{eqn:z-transverse}
\mathbf{V}_{i,j,k,E,W}^{n+1/2}=
\mathbf{V}_{i,j,k,E,W}^{n+1/2,y}
-\frac{\Delta t}{2\Delta z}\mathbf{A}_z(\mathbf{V}_{i,j,k}^n)\Delta_{z}^{up}\mathbf{V}_{i,j,k}^{n}.
\end{equation}
In these transverse
corrector steps, it is important to use the {\it{upwind}} biased slope limiter
instead of  any form of TVD limiters as reviewed in Section \ref{sec:CharacteristicTracingForTransverse}. Note that in the original 2D CTU
scheme by Colella [\citenum{Colella1990}], using the upwind flux gradients
in the transverse directions is the key mechanism that guarantees the full
CFL stability bound. We
establish the same upwind couplings by means of using the upwind slope limiter for
our transverse corrector. Using a TVD slope limiter instead would result in a
reduced stability limit for our algorithm (and we avoid using it).
The two transverse correction terms in (\ref{eqn:y-transverse}) and (\ref{eqn:z-transverse}) are calculated as in Section \ref{sec:CharacteristicTracingForTransverse},  
completing our description of the reduced 3D CTU scheme.

\subsubsection{Full 3D CTU Scheme in USM: Interface State Update from $n$ to $n+1/2$ Time Step}
\label{sec:full3dCTU-usm}
To establish the full stability limit (CFL number less than 1 in 3D) as
featured in the 12-solve CTU scheme of Saltzman [\citenum{Saltzman1994}],
we need one more step to couple diagonally moving flow effects. This
situation occurs when the conservative quantities are advected across the
corners diagonally with components of the local velocity fields $(u,v,w)$ being of
comparable orders of magnitude. In USM, these couplings can be
added to the interface states by performing intermediate state calculations at $n+\frac{1}{3}$. 
They involve
extra evaluations of 
the coefficient matrices and
the undivided upwind differences in (\ref{eqn:y-transverse}) and
(\ref{eqn:z-transverse}) at
\begin{equation}
\mathbf {{V}}_{i,j,k}^{n+{1}/{3},z}=
\mathbf {{V}}_{i,j,k}^{n}
-\frac{\Delta t}{3\Delta z}(\mathbf {{A}}_z)_{i,j,k}^n \Delta_{z}^{up}\mathbf{V}_{i,j,k}^{n},
\label{sec2:TransversalCorrector-z_fullCTU}
\end{equation}
and 
\begin{equation}
\mathbf {{V}}_{i,j,k}^{n+{1}/{3},y}=
\mathbf {{V}}_{i,j,k}^{n}
-\frac{\Delta t}{3\Delta y}(\mathbf {{A}}_y)_{i,j,k}^n \Delta_{y}^{up}\mathbf{V}_{i,j,k}^{n}.
\label{sec2:TransversalCorrector-y_fullCTU}
\end{equation}
More specifically, the transverse correctors in Equations 
(\ref{eqn:y-transverse}) and (\ref{eqn:z-transverse}) are replaced by
\begin{equation}
\label{eqn:y-transverse-full}
\mathbf{V}_{i,j,k,E,W}^{n+1/2,y}=
\mathbf{V}_{i,j,k,E,W}^{n+1/2,\parallel}
-\frac{\Delta t}{2\Delta y}\mathbf{A}_y(\mathbf{V}_{i,j,k}^{n+1/3,z}) \Delta_{y}^{up}\mathbf{V}_{i,j,k}^{n+1/3,z},
\end{equation}
and
\begin{equation}
\label{eqn:z-transverse-full}
\mathbf{V}_{i,j,k,E,W}^{n+1/2}=
\mathbf{V}_{i,j,k,E,W}^{n+1/2,y}
-\frac{\Delta t}{2\Delta z}\mathbf{A}_z(\mathbf{V}_{i,j,k}^{n+1/3,y})  \Delta_{z}^{up}\mathbf{V}_{i,j,k}^{n+1/3,y}.
\end{equation}

Here we make one important observation. Note that the additional
re-evaluations of the matrices $\mathbf{A}_y$ and $\mathbf{A}_z$ at the
$n+\frac{1}{3}$ states $\mathbf {{V}}_{i,j,k}^{n+{1}/{3},z}$ and $\mathbf
{{V}}_{i,j,k}^{n+{1}/{3},y}$ simply mean that the corresponding
eigensystems for the characteristic tracing in the transverse directions
need to be re-calculated, incurring the corresponding additional cost. Considering the full 3D
interface state calculations in Equations  (\ref{sec2:Evolx_2})-(\ref{sec2:Evolz_2}),
there are a total of six additional eigensystem evaluations required for the
transverse correctors, which becomes as expensive as directly solving
the corresponding Riemann problems, making our scheme expensive. Therefore
an efficient alternative approach is required.
Noticing
\begin{equation}
\frac{\Delta t}{3\Delta z}(\mathbf {{A}}_z)_{i,j,k}^n {\Delta}_{z}^{up}\mathbf{V}_{i,j,k}^n
=\frac{\Delta t}{3}\frac{\partial \mathbf{H}}{\partial z}\Biggl|_{\mathbf{V}_{i,j,k}^{n}},
\end{equation}
and using a Taylor expansion at $\mathbf{V}_{i,j,k}^{n}$, we consider
\begin{equation}
\label{eqn:first-order-matrixEvaluation}
\mathbf{A}_y(\mathbf{V}_{i,j,k}^{n+1/3,z})
=\frac{\partial \mathbf{G}}{\partial \mathbf{V}}\Biggl|_{\mathbf{V}_{i,j,k}^{n+1/3,z}}
=\frac{\partial \mathbf{G}}{\partial \mathbf{V}}\Biggl|_{\mathbf{V}_{i,j,k}^{n}}
-\frac{\Delta t}{3}\frac{\partial \mathbf{H}}{\partial z}\Biggl|_{\mathbf{V}_{i,j,k}^{n}}
\frac{\partial^2 \mathbf{G}}{\partial \mathbf{V}^2}\Biggl|_{\mathbf{V}_{i,j,k}^{n}}
=\mathbf{A}_y(\mathbf{V}_{i,j,k}^{n}) + \mathcal{O}(\Delta t).
\end{equation}
Ignoring the $\Delta t$ error term in the matrix evaluations in Equation (\ref{eqn:first-order-matrixEvaluation}), 
we can replace respectively $\mathbf{A}_y(\mathbf{V}_{i,j,k}^{n+1/3,z})$ and $\mathbf{A}_z(\mathbf{V}_{i,j,k}^{n+1/3,y})$ with
$\mathbf{A}_y(\mathbf{V}_{i,j,k}^{n})$ and $\mathbf{A}_z(\mathbf{V}_{i,j,k}^{n})$ 
in Equations  (\ref{eqn:y-transverse-full})-(\ref{eqn:z-transverse-full}).
However, it is essential to retain
$\Delta_{y}^{up}\mathbf{V}_{i,j,k}^{n+1/3,z}=\biggl({\partial \mathbf{V}}\Bigl/{\partial y}\Bigl|_{\mathbf{V}_{i,j,k}^{n+1/3,z}}\biggr)  \Delta y$ 
and
$\Delta_{z}^{up}\mathbf{V}_{i,j,k}^{n+1/3,y}=\biggl({\partial \mathbf{V}}\Big/{\partial z}\Bigl|_{\mathbf{V}_{i,j,k}^{n+1/3,y}}\biggr) \Delta z$ 
to couple the diagonal upwind
corner transport. We proceed this as follows. 
Ignoring the $\mathcal{O}(\Delta t)$ term and keeping the first-order
approximation in Equation (\ref{eqn:first-order-matrixEvaluation}) for the matrix
evaluation, 
the transverse corrector in Equation (\ref{eqn:y-transverse-full})  becomes
\begin{equation}
\label{eqn:y-transverse-full-approx}
\mathbf{V}_{i,j,k,E,W}^{n+1/2,y}=
\mathbf{V}_{i,j,k,E,W}^{n+1/2,\parallel}
-\frac{\Delta t}{2\Delta y}\mathbf{A}_y(\mathbf{V}_{i,j,k}^{n})\Delta_{y}^{up}\mathbf{V}_{i,j,k}^{n+1/3,z}
\end{equation}
Using our transverse corrector strategy, we get
\begin{equation}
\label{sec2:Transverse-y-EW-approx}
\mathbf {\hat {V}}_{y,i,j,k,E,W}^{n+1/2,y}=
\mathbf {\hat {V}}_{y,i,j,k,E,W}^{n+1/2,\parallel}-
\frac{\Delta t}{2\Delta y}
\sum_{m=1}^{7}\lambda^m_{y,i,j,k}\mathbf{r}^m_{y,i,j,k}
{\Delta}_{y}^{up}\hat\alpha_{i,j,k}^{n+1/3,z} 
-\frac{\Delta t}{2\Delta y}(\mathbf {A}_{B_y})_{i,j,k}^n 
{\Delta} B^{n+1/3,z}_{y,j},
\end{equation}
where the upwinding slope applied to each characteristic variable $\hat\alpha$ is given by
\begin{eqnarray}
\label{sec2:UpwindingLimiterCharY-approx}
{\Delta}_{y}^{up}\hat\alpha_{i,j,k}^{n+1/3,z} &=&
\cases{\mathbf{l}^m_{y,i,j,k}\cdot (\hat{\mathbf{V}}_{i,j+1,k}^{n+1/3,z}-\hat{\mathbf{V}}_{i,j,k}^{n+1/3,z}) \;\;\;\; \mbox{if} \;\;\; \lambda^m_{y,i,j,k} < 0\cr
             \mathbf{l}^m_{y,i,j,k}\cdot (\hat{\mathbf{V}}_{i,j,k}^{n+1/3,z}-\hat{\mathbf{V}}_{i,j-1,k}^{n+1/3,z}) \;\;\;\; \mbox{if} \;\;\; \lambda^m_{y,i,j,k} >0}.
\end{eqnarray}
Notice that the MHD term at $n+\frac{1}{3}$ can be written as
\begin{equation}
{\Delta} B^{n+1/3,z}_{y,j}={\Delta}_{y} 
\biggl( B_{y,j}^n - \frac{\Delta t}{3 \Delta z} 
\bigl[(\hat{\mathbf{{A}}}_z)^{n}_{i,j,k} {\Delta}_{z}^{up}\hat\mathbf{V}_{i,j,k}^{n}+
({\mathbf{{A}}}_{B_z})^{n}_{i,j,k}     {\Delta} B^n_{z,k}\bigr] \cdot \mathbf {e}_{B_y}\biggr)
={\Delta}_{y} 
\biggl( B_{y,j}^n + \mathcal{O}(\Delta t)\biggr),
\end{equation}
where $\mathbf {e}_{B_y}$ is a unit vector in $B_y$ direction for contraction and the hat notation implies the 
omission of the $B_z$ components. However, in order to choose ${\Delta} B^{n+1/3,z}_{y,j}$ to
enforce the numerical divergence 
to be zero always (see Equation (\ref{sec2:DivFreeConstraint})), we further 
drop the $\Delta t$ error term and only take
\begin{equation}
{\Delta} B^{n+1/3,z}_{y,j}={\Delta}_{y} B_{y,j}^n = b^n_{y,j+1/2}-b^n_{y,j-1/2},
\end{equation}
where $b^n_{y,j\pm1/2}$ are the cell face-centered, divergence-free magnetic fields in $y$-direction.

The upwind differences in
relation (\ref{sec2:UpwindingLimiterCharY-approx}) are given by (assuming uniform
spacing in each direction everywhere),
\begin{eqnarray}
\label{eqn:delta_z-plus-approx-3dcorrect}
&&{\mathbf{\hat{V}}}^{n+1/3,z}_{i,j+1,k}-{\mathbf{\hat{V}}}^{n+1/3,z}_{i,j,k} \nonumber \\
&&={\mathbf{\hat{V}}}^{n}_{i,j+1,k}-{\mathbf{\hat{V}}}^{n}_{i,j,k} -
\frac{\Delta t}{3\Delta z}
\Bigl[
(\hat{\mathbf{{A}}}_z)_{i,j+1,k}^n{{\Delta}}_{z}^{up}\hat\mathbf{V}_{i,j+1,k}^n + ({\mathbf{{A}}}_{B_z}^{n}{\Delta} B^n_{z})_{i,j+1,k} -
(\hat{\mathbf{{A}}}_z)_{i,j,  k}^n{{\Delta}}_{z}^{up}\hat\mathbf{V}_{i,j,k}^n - ({\mathbf{{A}}}_{B_z}^{n}{\Delta} B^n_{z})_{i,j,  k}
\Bigr] \nonumber \\
&&={\mathbf{\hat{V}}}^{n}_{i,j+1,k}-{\mathbf{\hat{V}}}^{n}_{i,j,k} -\frac{\Delta t}{3\Delta z}
\Bigl[ 
 \sum_{h=1}^{7}\lambda^{h}_{z,i,j+1,k}\mathbf{r}^{h}_{z,i,j+1,k} {\Delta}_z^{up}\hat\alpha_{i,j+1,k}^{n}
-\sum_{l=1}^{7}\lambda^{l}_{z,i,j,  k}\mathbf{r}^{l}_{z,i,j,  k}  {\Delta}_z^{up}\hat\alpha_{i,j,k}^{n}
\nonumber \\ 
&&+
({\mathbf{{A}}}_{B_z}^{n}{\Delta} B^n_{z})_{i,j+1,k}-({\mathbf{{A}}}_{B_z}^{n}{\Delta} B^n_{z})_{i,j,  k}
\Bigr], 
\end{eqnarray}
and
\begin{eqnarray}
\label{eqn:delta_z-minus-approx-3dcorrect}
&&{\mathbf{\hat{V}}}^{n+1/3,z}_{i,j,k}-{\mathbf{\hat{V}}}^{n+1/3,z}_{i,j-1,k} \nonumber \\
&&={\mathbf{\hat{V}}}^{n}_{i,j,k}-{\mathbf{\hat{V}}}^{n}_{i,j-1,k} -
\frac{\Delta t}{3\Delta z}
\Bigl[
(\hat{\mathbf{{A}}}_z)_{i,j,k}^n{{\Delta}}_{z}^{up}\hat\mathbf{V}_{i,j,k}^n + ({\mathbf{{A}}}_{B_z}^{n}{\Delta} B^n_{z})_{i,j,k} -
(\hat{\mathbf{{A}}}_z)_{i,j-1,k}^n{{\Delta}}_{z}^{up}\hat\mathbf{V}_{i,j-1,k}^n - ({\mathbf{{A}}}_{B_z}^{n}{\Delta} B^n_{z})_{i,j-1,  k}
\Bigr] \nonumber \\
&&={\mathbf{\hat{V}}}^{n}_{i,j,k}-{\mathbf{\hat{V}}}^{n}_{i,j-1,k} -\frac{\Delta t}{3\Delta z}
\Bigl[ 
 \sum_{h=1}^{7}\lambda^{h}_{z,i,j,k}\mathbf{r}^{h}_{z,i,j,k} {\Delta}_z^{up}\hat\alpha_{i,j,k}^{n}
-\sum_{l=1}^{7}\lambda^{l}_{z,i,j-1,  k}\mathbf{r}^{l}_{z,i,j-1,  k}  {\Delta}_z^{up}\hat\alpha_{i,j-1,k}^{n}
\nonumber \\ 
&&+
({\mathbf{{A}}}_{B_z}^{n}{\Delta} B^n_{z})_{i,j,k}-({\mathbf{{A}}}_{B_z}^{n}{\Delta} B^n_{z})_{i,j-1,  k}
\Bigr], 
\end{eqnarray}
In the case of $\lambda^m_{y,i,j,k}>0$ for all $m$, 
using (\ref{eqn:delta_z-minus-approx-3dcorrect}), Equation (\ref{sec2:Transverse-y-EW-approx}) becomes
\begin{eqnarray}
\label{sec2:Transverse-y-EW-approx-3dcorrect_a}
&&\mathbf {\hat {V}}_{y,i,j,k,E,W}^{n+1/2,y}
=\mathbf {\hat {V}}_{y,i,j,k,E,W}^{n+1/2,\parallel}
-\frac{\Delta t}{2\Delta y}
\sum_{m=1}^{7}\lambda^m_{y,i,j,k}\mathbf{r}^m_{y,i,j,k}
\mathbf{l}^m_{y,i,j,k}\cdot ({\mathbf{\hat{V}}}^{n+1/3,z}_{i,j,k}-{\mathbf{\hat{V}}}^{n+1/3,z}_{i,j-1,k})
-\frac{\Delta t}{2\Delta y}(\mathbf {A}_{B_y})_{i,j,k}^n \Delta B^{n}_{y,j} \\
\label{sec2:Transverse-y-EW-approx-3dcorrect_b}
&&=\mathbf {\hat {V}}_{y,i,j,k,E,W}^{n+1/2,\parallel}
-\frac{\Delta t}{2\Delta y}
\sum_{m=1}^{7}\lambda^m_{y,i,j,k}\mathbf{r}^m_{y,i,j,k}
\mathbf{l}^m_{y,i,j,k}\cdot 
({\mathbf{\hat{V}}}^{n}_{i,j,k}-{\mathbf{\hat{V}}}^{n}_{i,j-1,k})
-\frac{\Delta t}{2\Delta y}(\mathbf {A}_{B_y})_{i,j,k}^n \Delta B^{n}_{y,j} \\ 
\label{sec2:Transverse-y-EW-approx-3dcorrect_c}
&&+\frac{\Delta t^2}{6\Delta y \Delta z}
\Biggl(\sum_{m=1}^{7}\lambda^m_{y,i,j,k}\mathbf{r}^m_{y,i,j,k}\mathbf{l}^m_{y,i,j,k}\cdot 
\Bigl[\sum_{h=1}^{7}\lambda^h_{z,i,j,k}\mathbf{r}^h_{z,i,j,k} {\Delta}_z^{up}\hat\alpha_{i,j,k}^{n}
     -\sum_{l=1}^{7}\lambda^l_{z,i,j-1,k}\mathbf{r}^l_{z,i,j-1,k} {\Delta}_z^{up}\hat\alpha_{i,j-1,k}^{n}
\nonumber \\
&&+
({\mathbf{{A}}}_{B_z}^{n}{\Delta} B^n_{z})_{i,j,k}-({\mathbf{{A}}}_{B_z}^{n}{\Delta} B^n_{z})_{i,j-1,  k}
\Bigr]
\Biggr).
\end{eqnarray}
Note that the terms in relation (\ref{sec2:Transverse-y-EW-approx-3dcorrect_b}) are what we already
established in the reduced 3D CTU scheme. 
The terms in 
relation (\ref{sec2:Transverse-y-EW-approx-3dcorrect_c}) are new correction terms for
the full 3D CTU scheme that need to be added to the reduced 3D CTU
interface states in relation (\ref{sec2:Transverse-y-EW-approx-3dcorrect_b}).

Similarly, the final form of $x$-interface states
$\mathbf{V}_{i,j,k,E,W}^{n+1/2}$ in Equation (\ref{eqn:z-transverse-full}) is
established by adding another correction term
appearing in 
${\Delta}_{z}^{up}\hat\mathbf{V}_{i,j,k}^{n+1/3,y}$, 
giving the result (assuming $\lambda^m_{z,i,j,k}>0$ 
for all $m$)
\begin{eqnarray}
\label{sec2:Transverse-z-EW-approx-3dcorrect_a}
&&\mathbf{V}_{i,j,k,E,W}^{n+1/2}=\mathbf {\hat {V}}_{y,i,j,k,E,W}^{n+1/2,y}
-\frac{\Delta t}{2\Delta z}
\sum_{m=1}^{7}\lambda^m_{z,i,j,k}\mathbf{r}^m_{z,i,j,k}
\mathbf{l}^m_{z,i,j,k}\cdot ({\mathbf{\hat{V}}}^{n}_{i,j,k}-{\mathbf{\hat{V}}}^{n}_{i,j,k-1})
-\frac{\Delta t}{2\Delta z}(\mathbf {A}_{B_z})_{i,j,k}^n \Delta B^n_{z,k} \\
\label{sec2:Transverse-z-EW-approx-3dcorrect_b}
&&+\frac{\Delta t^2}{6\Delta z \Delta y}
\Biggl(\sum_{m=1}^{7}\lambda^m_{z,i,j,k}\mathbf{r}^m_{z,i,j,k}\mathbf{l}^m_{z,i,j,k}\cdot 
\Bigl[\sum_{h=1}^{7}\lambda^h_{y,i,j,k  }\mathbf{r}^h_{y,i,j,k  } {\Delta}_y^{up}\hat\alpha_{i,j,k  }^{n}
     -\sum_{l=1}^{7}\lambda^l_{y,i,j,k-1}\mathbf{r}^l_{y,i,j,k-1} {\Delta}_y^{up}\hat\alpha_{i,j,k-1}^{n}
\nonumber \\
&&+
({\mathbf{{A}}}_{B_y}^{n}{\Delta} B^n_{y})_{i,j,k}-({\mathbf{{A}}}_{B_y}^{n}{\Delta} B^n_{y})_{i,j,  k-1}
\Bigr]
\Biggr).
\end{eqnarray}
Likewise, the terms in relation (\ref{sec2:Transverse-z-EW-approx-3dcorrect_b}) are
the extra correction terms required for the full 3D CTU scheme. They must be
added to the reduced CTU terms in relation (\ref{sec2:Transverse-z-EW-approx-3dcorrect_a}).

It is worth pointing out at this stage
that all of the eigensystems in the full 3D CTU
correction terms are readily available as they have been calculated and
stored in the normal predictor step in each $x,y,z$-direction described in Section
\ref{sec:normal-predictor-dataReconst}. In the normal predictor step, one
can store not only the eigensystems, but also the two summations inside the
square brackets in relations (\ref{sec2:Transverse-y-EW-approx-3dcorrect_c}) and
(\ref{sec2:Transverse-z-EW-approx-3dcorrect_b}) (see also the simple pseudo code
in Section \ref{sec:CharacteristicTracingForTransverse}). The only extra
calculations imposing additional computational costs are therefore the
upwind differencings inside the square brackets and the dot products in
relations (\ref{sec2:Transverse-y-EW-approx-3dcorrect_c}) and
(\ref{sec2:Transverse-z-EW-approx-3dcorrect_b}),  which are computationally
much more efficient compared to the calculation requirements in the
12-solve CTU scheme.
This completes our description of the single-step, data
reconstruction-evolution algorithm for all variables {\em except} the
divergence-free normal magnetic fields at each cell face.
The reconstructed interface states are second-order accurate in space and evolved to
$n+\frac{1}{2}$ time step at each interface. The next step is to advance the
remaining normal fields at the cell faces, finalizing the Riemann
state calculations.

\subsection{Advancing the Normal Fields from $n$ to $n+1/2$ Time Step using CT}
\label{sec:CThalfTimeStepUpdate}
In updating the normal fields to $n+\frac{1}{2}$, it is important to meet two conditions.
The first is a continuity restriction of the normal magnetic field across
cell interfaces 
[\citenum{Powell1994,Balsara2003,Crockett2005,GardinerStone2005,LeeDeane2009}].
The second is the divergence-free constraint of the normal fields on a
computational grid.
As a last step of our Riemann state calculations, we must evolve the normal
field components at each cell boundary by a half time step, while satisfying the
two conditions. We therefore follow the CT approach using the
high-order Godunov fluxes that are solutions to a Riemann problem using the
Riemann states $\mathbf V^{n+1/2}_{i,j,N,S,E,W,T,B}$ described in Sections
\ref{sec:reduced3dCTU-usm} and \ref{sec:full3dCTU-usm}. Our approach here
is the 3D extension of the 2D method in using the same approach as in
[\citenum{LeeDeane2009}].
Here we briefly describe the procedure only in $x$-direction, which can be similarly applied to
the other directions.
We first solve Riemann problem at $x$ interfaces as
\begin{equation}
\tilde{\mathbf {F}}_{i-1/2,j,k}^{*,n+1/2} =
\mbox{RP}\left(\mathbf {{V}}_{i-1,j,k,E}^{n+1/2},\mathbf { {V}}_{i,j,k,W}^{n+1/2} \right), \;\;\;\;
\tilde{\mathbf {F}}_{i+1/2,j,k}^{*,n+1/2} =
\mbox{RP}\left(\mathbf {{V}}_{i,j,k,E}^{n+1/2},\mathbf { {V}}_{i+1,j,k,W}^{n+1/2} \right),
\label{sec2:RP_F_Bnormal}
\end{equation}
With these high-order Godunov fluxes at the half time step we evolve the normal fields by a half time step using the CT update 
\begin{eqnarray}
{b}^{n+1/2}_{x,i+1/2,j,k}&&=
{b}^{n}_{x,i+1/2,j,k}\nonumber\\
&&-\frac{\Delta t}{2\Delta y}
\Bigr\{
 \tilde{E}^{*,n+1/2}_{z,i+1/2,j+1/2,k} 
-\tilde{E}^{*,n+1/2}_{z,i+1/2,j-1/2,k}
\Bigl\}
-\frac{\Delta t}{2\Delta z}
\Bigr\{
-\tilde{E}^{*,n+1/2}_{y,i+1/2,j,k+1/2} 
+\tilde{E}^{*,n+1/2}_{y,i+1/2,j,k-1/2}
\Bigl\},
\label{sec2:InductionBx2D}
\end{eqnarray}
where the duality relationship between the electric fields and the
high-order Godunov fluxes [\citenum{BalsaraSpicer1999}] 
is assumed in the expression.
The electric fields $\tilde{E}_z^{*,n+1/2}$
\footnote{Note here that we use a consistent superscript (e.g., $\tilde \mathbf{F}^*$ and $\tilde E^*$) between the Godunov fluxes and the electric fields that are in the duality relationship. The superscript  is used for the intermediate Riemann solutions in Section \ref{sec:CThalfTimeStepUpdate}, whereas the superscript $*$ (e.g., $\mathbf{F}^*$ and $E^*$) is used for the final Riemann solutions in Section \ref{sec:unkUpdate}.} 
in (\ref{sec2:InductionBx2D})
can be constructed based on the MEC method [\citenum{LeeDeane2009}] that takes an arithmetic average of four Taylor series expansions of the fluxes given by
(\ref{sec2:RP_F_Bnormal})
to obtain them (e.g., see Equation (\ref{sec2:TaylorEz}) in Section \ref{sec:standardMEC}).
The normal fields in (\ref{sec2:InductionBx2D})
satisfy the divergence-free constraint as well as the
continuity restriction across cell interfaces as they are direct solutions
to numerical induction equations via the CT approach. 

Given these updated cell face-centered divergence-free fields, the Riemann states at $x$-interfaces are updated as
\begin{eqnarray}
\label{sec2:UpdateInterfaceX}
&&\mathbf V^{n+1/2}_{i,j,k,E} \cdot \mathbf {e}_{B_x} =b^{n+1/2}_{x,i+1/2,j,k},\;\;\;\;
     \mathbf V^{n+1/2}_{i,j,k,W} \cdot \mathbf {e}_{B_x}=b^{n+1/2}_{x,i-1/2,j,k},
\label{sec2:UpdateInterfaceY}
\end{eqnarray}
where $\mathbf {e}_{B_x}$ are unit vectors for the magnetic field components in 
$x$-direction.

Now that the second-order accurate Riemann states $\mathbf {
{V}}_{i,j,k,N,S,E,W,T,B}^{n+1/2}$ are available, the second-order Godunov
fluxes can be evaluated by solving the last set of 
Riemann problems
at $x$-interfaces,
\begin{equation}
\mathbf {F}_{i-1/2,j,k}^{*,n+1/2} =
\mbox{RP}\left(\mathbf {{V}}_{i-1,j,k,E}^{n+1/2},\mathbf { {V}}_{i,j,k,W}^{n+1/2} \right), \;\;\;\;
\mathbf {F}_{i+1/2,j,k}^{*,n+1/2} =
\mbox{RP}\left(\mathbf {{V}}_{i,j,k,E}^{n+1/2},\mathbf { {V}}_{i+1,j,k,W}^{n+1/2} \right),
\label{sec2:RP_F}
\end{equation}
Note that the superscript $*$ is used to represent the second-order Godunov fluxes that are the solutions of the Riemann problems.

\subsection{Cell-centered Solution Update from $n$ to $n+1$ Time Step}
\label{sec:unkUpdate}
The USM algorithm updates the cell-centered conservative variables to the next time step $n+1$ using an unsplit integrator,
\begin{equation}
\mathbf {U}_{i,j,k}^{n+1}=\mathbf {U}_{i,j,k}^{n} 
\frac{\Delta t}{{\Delta x}}
\left\{{\mathbf {F}_{i+1/2,j,k}^{*,n+1/2} 
        -\mathbf {F}_{i-1/2,j,k}^{*,n+1/2}}\right\}-
\frac{\Delta t}{{\Delta y}}
\left\{{\mathbf {G}_{i,j+1/2,k}^{*,n+1/2}
        -\mathbf {G}_{i,j-1/2,k}^{*,n+1/2}}\right\}
        -\frac{\Delta t}{{\Delta z}}
\left\{{\mathbf {H}_{i,j,k+1/2}^{*,n+1/2}
         -\mathbf {H}_{i,j,k-1/2}^{*,n+1/2}}\right\}.        
\label{sec2:unsplit_update}
\end{equation}
In general, after this update, non-zero divergence magnetic fields are still present at cell centers, 
and they need to be corrected. 
In the next section we update the divergence-free cell face-centered
magnetic fields from $n$ to $n+1$ time step using the
modified electric field construction (MEC) scheme [\citenum{LeeDeane2009}].
The cell face-centered fields are averaged to correct the cell-centered magnetic fields at the $n+1$ state.
The choice of a time step $\Delta t$ for the full 3D CTU scheme is limited by the full CFL bound, 
with which we set our CFL number to be $0.95$ for all numerical results presented in this paper, unless otherwise stated.

\subsection{Face-centered, Divergence-Free Fields Update via CT from $n$ to $n+1$ Time Step using MEC}
\label{sec:MEC}
\subsubsection{The Standard Arithmetic Averaging Approach in MEC: standard-MEC}
\label{sec:standardMEC}
In [\citenum{LeeDeane2009}], the 2D version of the modified electric field
construction (MEC) scheme was introduced. 
The method provides 
electric fields at cell corners using high-order Taylor
expansions. Displaying the electric field in $z$-direction only, 
this standard-MEC algorithm gives 

\begin{eqnarray}
\cases{
E^{  n+1/2}_{z,i+1/2,j+1/2,k}=
E^{*,n+1/2}_{z,i+1/2,j,    k} 
+\frac{\Delta y}{2}  \frac{\partial   E^{*,n+1/2}_{z,i+1/2,j,    k} }{\partial y}
+\frac{\Delta y^2}{8}\frac{\partial^2 E^{*,n+1/2}_{z,i+1/2,j,    k} }{\partial y^2}
+\mathcal{O}(\Delta y^3),
\cr
E^{  n+1/2}_{z,i+1/2,j+1/2,k}=
E^{*,n+1/2}_{z,i+1/2,j+1,  k}
-\frac{\Delta y}{2}  \frac{\partial   E^{*,n+1/2}_{z,i+1/2,j+1,  k}}{\partial y}
+\frac{\Delta y^2}{8}\frac{\partial^2 E^{*,n+1/2}_{z,i+1/2,j+1,  k}}{\partial y^2}
+\mathcal{O}(\Delta y^3),
\cr
E^{  n+1/2}_{z,i+1/2,j+1/2,k}=
E^{*,n+1/2}_{z,i,    j+1/2,k}
+\frac{\Delta x}{2}  \frac{\partial   E^{*,n+1/2}_{z,i,    j+1/2,k}}{\partial x}
+\frac{\Delta x^2}{8}\frac{\partial^2 E^{*,n+1/2}_{z,i,    j+1/2,k}}{\partial x^2}
+\mathcal{O}(\Delta x^3),
\cr
E^{  n+1/2}_{z,i+1/2,j+1/2,k}=
E^{*,n+1/2}_{z,i+1,  j+1/2,k}
-\frac{\Delta x}{2}  \frac{\partial   E^{*,n+1/2}_{z,i+1,  j+1/2,k}}{\partial x}
+\frac{\Delta x^2}{8}\frac{\partial^2 E^{*,n+1/2}_{z,i+1,  j+1/2,k}}{\partial x^2}
+\mathcal{O}(\Delta x^3).
\cr}
\label{sec2:TaylorEz}
\end{eqnarray}
The duality relationship [\citenum{BalsaraSpicer1999}] has been assumed
for those electric fields at cell face centers about which the Taylor
series are expanded.
The standard-MEC algorithm proceeds to take a simple arithmetic average of these four Taylor
expansions of each field in Equation 
(\ref{sec2:TaylorEz}) to
get an averaged electric field 
$\tilde{E}^{n+1/2}_{z,i+1/2,j+1/2,k}$.

\subsubsection{Upwind Biased Averaging in MEC: upwind-MEC}
\label{sec:upwindMEC}
The standard CT approach of taking the arithmetic average of the four electric fields 
is simple enough to work well in local smooth regions. The simplest form of this averaging approach was first suggested by Balsara and Spicer [\citenum{BalsaraSpicer1999}] \noter{using 1D based Riemann solvers.} 
This idea seems a very natural choice if one considers the grid locations of the electric fields. However, the authors understood that in truly multidimensional flows where there is a directional bias, the simple arithmetic averaging scheme may need to be corrected and it would be better to incorporate upwind information. \noter{Recently, a general resolution on such issue with multidimensional upwinding has become available by the subsequent efforts to build the genuinely multidimensional HLL-type Riemann solvers by Balsara [\citenum{Balsara2010, Balsara2012}].}

On the other hand, within the 1D Riemann based CTU approach, Gardiner and Stone [\citenum{GardinerStone2005}] identified the shortcomings of the simple arithmetic averaging method and developed a systematic construction of CT algorithms that are consistent for a plane-parallel, grid-aligned flow. They recovered the necessary amount of numerical viscosity that stabilizes their underlying CT integration algorithm.
The CT methods proposed therein readily satisfy 
planar symmetry for $\partial/{\partial x}=0$ or  $\partial/{\partial y}=0$, showing that their algorithms recover the associated one-dimensional solution for the underlying integration algorithm. \noter{A similar approach of increasing dissipation is also found in [\citenum{LondrilloDelZanna2004}].}

Although the approach by Gardiner and Stone provides consistency for plane-parallel, grid-aligned flows, the method does not take into account multidimensional effects where the flow has one specific directional bias without assuming $\partial/{\partial x}=0$ or  $\partial/{\partial y}=0$. 
To illustrate this, we consider the weakly magnetized field loop advection problem [\citenum{GardinerStone2005,LeeDeane2009}] where the loop is advected by a dominant velocity in $x$ and comparably small velocity in $y$, saying $u>0$ with $v=\epsilon>0$. 
The simple arithmetic averaging CT algorithm gives 
\begin{equation}
\tilde E^{  n+1/2}_{z,i+1/2,j+1/2,k}=\frac{1}{4}\bigl(E^{*,n+1/2}_{z,i+1/2,j,k} + E^{*,n+1/2}_{z,i+1/2,j+1,k} +  E^{*,n+1/2}_{z,i,j+1/2,k} + E^{*,n+1/2}_{z,i+1,j+1/2,k}\bigr).
\end{equation}
In the limiting case of $u>0$ with $v=\epsilon \rightarrow 0$, it is obvious that $E^{*,n+1/2}_{z,i,j+1/2,k}$ is the only electric field that is in the upwind direction, whereas the rest are not. 
\noter{One can easily see that the similar situation also occurs in the CT scheme in [\citenum{GardinerStone2005}].}
This suggests that the simple averaging, \noter{based on the 1D Riemann solver strategies used with CTU,}
is potentially exposed to numerical oscillations and therefore its stability is questionable.
\noter{There are several other modern time-evolution strategies for MHD that do not suffer from this lack of upwinding. 
As noted, the modern MHD schemes by Balsara [\citenum{Balsara2010, Balsara2012}] using 
the genuinely multidimensional Riemann solvers evolve the magnetic field structures in any direction without resorting
to any added dissipation in the electric fields. The use of multidimensional Riemann solvers for MHD have shown to possess superior capability
in evolving magnetic fields to the use of conventional 1D Riemann solvers, better reflecting the true nature of the PDE that does not
require any secondary dissipation mechanisms for the purpose of stable upwinding.
Although the essential role of the multidimensional technology is 
acknowledged, our primary goal in this paper is to design an easy alternative that can be ameliorated within the 1D Riemann solver framework based on CTU.}

We now describe our new upwind CT construction scheme that resolves this lack of upwind information in the current strategy.
As suggested, the idea is to construct the electric fields at $(i+\frac{1}{2},j+\frac{1}{2},k)$ including the electric fields at cell interfaces that are in the upwind directions. For example, in the limiting case of $u>0$ with $v=0$ the cell-cornered electric field is given by
\begin{equation}
\tilde E^{  n+1/2}_{z,i+1/2,j+1/2,k} = E^{*,n+1/2}_{z,i,j+1/2,k}.
\end{equation}

Based on this simple idea of upwinding, we illustrate a systematic approach to constructing a new upwind-MEC algorithm that also leads to a consistent CT scheme for plane-parallel, grid-aligned flows. To make our discussion more concise, we display a 2D case; 
the extension in 3D is straightforward. The first step is to check the upwind direction at each cell corner. This can be done by defining four switches
\begin{eqnarray}
u_P &=& \frac{1}{2}(1+\mbox{sign}(u_{i+1/2,j+1/2}))|\mbox{sign}(u_{i+1/2,j+1/2})|,\label{eqn:u_P}\\
u_N &=& \frac{1}{2}(1-\mbox{sign}(u_{i+1/2,j+1/2}))|\mbox{sign}(u_{i+1/2,j+1/2})|,\label{eqn:u_N}\\
v_P &=& \frac{1}{2}(1+\mbox{sign}(v_{i+1/2,j+1/2}))|\mbox{sign}(v_{i+1/2,j+1/2})|,\label{eqn:v_P}\\
v_N &=& \frac{1}{2}(1-\mbox{sign}(v_{i+1/2,j+1/2}))|\mbox{sign}(v_{i+1/2,j+1/2})|,\label{eqn:v_N}
\end{eqnarray}
where the sign function is defined by
\begin{eqnarray}
\mbox{sign}(x)=
\cases{1 \;\;\;\;\;\;\; \mbox{if} \;\;\; x>0,\cr
             0 \;\;\;\;\;\;\; \mbox{if} \;\;\; x=0,\cr
            -1 \;\;\;\; \mbox{if} \;\;\; x<0.}
\end{eqnarray}

The cell-centered $n$ time step velocity fields are spatially averaged to get the velocities at the cell corner $(i+\frac{1}{2},j+\frac{1}{2})$ in (\ref{eqn:u_P})--(\ref{eqn:v_N}).
When deciding the proper upwind direction at $(i+\frac{1}{2},j+\frac{1}{2})$ in Equations (\ref{eqn:u_P})-(\ref{eqn:v_N}), it is useful to measure relative magnitudes of velocity fields in order to avoid any numerical noise effects. The small noise perturbations in the signs of velocity fields may lead to an unnecessary amount of changes in upwind directions that are not very advantageous [\citenum{RumseyEtAl1993}].
This motivates to set the velocities at $(i+\frac{1}{2},j+\frac{1}{2})$ in (\ref{eqn:u_P})-(\ref{eqn:v_N}) to be zero whenever a given local velocity in one direction is relatively small compared to the total local velocity magnitude. That is to say, 
\begin{eqnarray}
u_{i+1/2,j+1/2}=0 \;\;\;\; \mbox{if} \;\;\;\;\; \frac{|u_{i+1/2,j+1/2}|}{\mbox{max}(\sqrt{u^2_{i+1/2,j+1/2}+v^2_{i+1/2,j+1/2}},\epsilon_2)} \le \epsilon_1.
\end{eqnarray}
Notice that the total velocity only includes the two velocity field components $u$ and $v$ (but {\it{not}} $w$) that define the electric field $E_z$ under consideration.
Our choice of \noter{an empirically derived value $\epsilon_1$  forces} to ignore any velocity fluctuations that are smaller than $10\%$ of the total magnitude of velocity fields, and set those velocities to be zero for determining the proper upwind direction. \noter{An arbitrary small value is chosen for $\epsilon_2$ to prevent division by zero.}

Finally we take an upwind biased averaging of the electric fields using the switches in (\ref{eqn:u_P})-(\ref{eqn:v_N}),
\begin{eqnarray}
\tilde E^{  n+1/2}_{z,i+1/2,j+1/2,k}=\alpha\Bigg[&&
v_P
\Biggl(
E^{*,n+1/2}_{z,i+1/2,j,    k} 
+\frac{\Delta y}{2}  \frac{\partial   E^{*,n+1/2}_{z,i+1/2,j,    k} }{\partial y}
+\frac{\Delta y^2}{8}\frac{\partial^2 E^{*,n+1/2}_{z,i+1/2,j,    k} }{\partial y^2}
\Biggr)+\nonumber \\
&&v_N
\Biggl(
E^{*,n+1/2}_{z,i+1/2,j+1,  k}
-\frac{\Delta y}{2}  \frac{\partial   E^{*,n+1/2}_{z,i+1/2,j+1,  k}}{\partial y}
+\frac{\Delta y^2}{8}\frac{\partial^2 E^{*,n+1/2}_{z,i+1/2,j+1,  k}}{\partial y^2}
\Biggr)+\nonumber \\
&&u_P
\Biggl(
E^{*,n+1/2}_{z,i,    j+1/2,k}
+\frac{\Delta x}{2}  \frac{\partial   E^{*,n+1/2}_{z,i,    j+1/2,k}}{\partial x}
+\frac{\Delta x^2}{8}\frac{\partial^2 E^{*,n+1/2}_{z,i,    j+1/2,k}}{\partial x^2}
\Biggr)+\nonumber \\
&&u_N
\Biggl(
E^{*,n+1/2}_{z,i+1,  j+1/2,k}
-\frac{\Delta x}{2}  \frac{\partial   E^{*,n+1/2}_{z,i+1,  j+1/2,k}}{\partial x}
+\frac{\Delta x^2}{8}\frac{\partial^2 E^{*,n+1/2}_{z,i+1,  j+1/2,k}}{\partial x^2}
\Biggr)
\Bigg].
\label{eqn:upwind-MEC}
\end{eqnarray}
Here the averaging weight factor $\alpha$ is set to 1 if $u_{i+1/2,j+1/2}v_{i+1/2,j+1/2}=0$; $\alpha=\frac{1}{2}$ otherwise. 
This is our upwind-MEC scheme.
It is interesting to observe that the upwind-MEC scheme satisfies the consistency relationship that reverts to the underlying integration CT scheme for plane-parallel, grid-aligned flows in an upwind sense. To see this we consider for example $\partial /\partial y = 0$. Consider first when $v_{i+1/2,j+1/2}=0$. In this case the electric fields at each cell corner will take only either the third (if $u_{i+1/2,j+1/2}>0$) or the fourth (if $u_{i+1/2,j+1/2}<0$) part of Equation (\ref{eqn:upwind-MEC}). By planar symmetry, $E^{  n+1/2}_{z,i,j+1/2,k}=E^{  n+1/2}_{z,i,j,k}=E^{  n+1/2}_{z,i,j+1,k}$, the first leading terms in both relationships in (\ref{eqn:upwind-MEC}) become
\begin{eqnarray}
\tilde E^{  n+1/2}_{z,i+1/2,j+1/2,k}=
\cases{E^{*, n+1/2}_{z,i,     j+1/2,k}=E^{*, n+1/2}_{z,i,    j,k} \;\;\;\;\;\;\; \mbox{if} \;\;\; u_{i+1/2,j+1/2}>0,\cr
             E^{*, n+1/2}_{z,i+1,j+1/2,k}=E^{*, n+1/2}_{z,i+1,j,k} \;\;\; \mbox{if} \;\;\; u_{i+1/2,j+1/2}<0.}
\label{eqn:upwind-MEC-plannarSymmetry}
\end{eqnarray}
Therefore they can be considered as an upwind-biased CT variant of $\tilde E^{  n+1/2}_{z,i+1/2,j+1/2,k}=E^{*, n+1/2}_{z,i+1/2,j,k}$
which is the result of the CT method by Gardiner and Stone [\citenum{GardinerStone2005}] in this case. If $u_{i+1/2,j+1/2}=0$ then $\tilde E^{  n+1/2}_{z,i+1/2,j+1/2,k}=0$ which is an exact solution for ideal MHD\footnote{Note that for non-ideal MHD the upwinding approach in the upwind-MEC scheme should only be applied to the induction part (i.e., $-{\mathbf{u}}\times {\mathbf{B}}$)  of a generalized Ohm's law including the terms such as the magnetic diffusion, the Hall effect and the Biermann battery effect.}.

For nonzero values of $v_{i+1/2,j+1/2}$, consider for instance a case for $v_{i+1/2,j+1/2}>0$ with $u_{i+1/2,j+1/2}>0$. Then the electric fields from the upwind-MEC scheme will take the parts that have $u_P$ and $v_P$ only, and $\alpha=\frac{1}{2}$. 
Consider only the first leading terms in the two parts of Equation (\ref{eqn:upwind-MEC}) for an exposition purpose, we get
\begin{equation}
\tilde E^{  n+1/2}_{z,i+1/2,j+1/2,k}=\frac{1}{2}\biggl(E^{*,n+1/2}_{z,i+1/2,  j,k}+E^{*,n+1/2}_{z,i,  j+1/2,k} \biggr)
=\frac{1}{2}\biggl(E^{*,n+1/2}_{z,i+1/2,  j+1/2,k}+E^{*,n+1/2}_{z,i,  j+1/2,k} \biggr),
\label{eqn:upwind-MEC-uv}
\end{equation}
where we assumed $\partial /\partial y = 0$ in the last equality. Compared to the electric field $\tilde E^{  n+1/2}_{z,i+1/2,j+1/2,k}=E^{ *, n+1/2}_{z,i+1/2,j,k}$ from the Gardiner and Stone's method, the upwind biased MEC electric field in Equation (\ref{eqn:upwind-MEC-uv}) makes use of an additional upwind electric field at $(i,j+\frac{1}{2},k)$ and includes that field in the average to get the field at $(i+\frac{1}{2},j+\frac{1}{2},k)$.

There are several important features of the upwind-MEC method. First, the method appropriately uses an upwinding direction rather than taking a simple arithmetic average which lacks proper upwinding.
The lack of upwinding is found in most of the well known CT schemes [\citenum{BalsaraSpicer1999,GardinerStone2005,GardinerStone2008,LeeDeane2009,MiniatiMartin2011}]. The upwinding strategy becomes most crucial when advecting a magnetized object in one biased direction, for instance, the weakly magnetized field loop advection problem [\citenum{GardinerStone2005}] in $x$-direction only or with a very small advection angle $\theta$ relative to $x$-axis. In this small angle advection case the standard CT update without any upwinding becomes very vulnerable to numerical instabilities that appear as spurious oscillations in the magnetic field evolution. Such oscillations are more likely in the small angle case than in a relatively large angle case because there is only one dominating direction from which the CT electric averaging scheme should rely on to obtain enough numerical dissipation to stabilize the field evolution. 
It will be shown later that the upwind-MEC strategy advects the field loop without significant numerical oscillations and without distortions for 
small angle advections. 

Second, the upwind-MEC scheme not only accounts for an upwinding direction for stability, but also includes high-order terms. 
The first derivative terms reflect correct spatial changes in expanding from the center nodes to the corners, while the second terms effectively avoid spurious oscillations near discontinuities by adding the proper amount of numerical dissipation to the corner extrapolated fields [\citenum{LeeDeane2009}]. These high-order terms are upwind averaged in such a way that the scheme is consistent for plane-parallel, grid-aligned flows.

Third, as mentioned, the idea of using upwinding in taking the average is to recover a {\it{proper}} amount of numerical dissipation required to ensure stability. We note that the greatest benefit occurs when there is a dominating direction locally towards which the magnetic fields are advected. For this reason the upwind-MEC scheme can be turned off when the local flow velocities are all ignorable. When the local velocities are all negligibly small but finite the local flow should be smooth enough, and hence it is sufficiently accurate to use the standard-MEC scheme that takes the arithmetic averaging as discussed in Section \ref{sec:standardMEC}. In practice, we switch back to the standard-MEC when the local flow velocities are relatively small compared to the local sound speed $C_s$. That is, we consider a local Mach number $M_z$ for the local flow switch to choose the standard-MEC for constructing the electric field $\tilde E_{z,i+1/2,j+1/2,k}$ if
\begin{equation}
M_z=\frac{\sqrt{u^2_{i+1/2,j+1/2}+v^2_{i+1/2,j+1/2}}}{C_s} \le \epsilon_3.
\end{equation}
\noter{An empirical based} tunable parameter $\epsilon_3=10^{-4}$ suffices to detect a local smooth flow in order to convert back to the standard-MEC method; otherwise the upwind-MEC scheme is enabled for all the numerical tests presented in this paper.  

\subsubsection{CT Update from $n$ to $n+1$ Time Step}
\label{sec:CTupdate}
Using the electric fields $\tilde{E}^{n+1/2}_{x,i,j+1/2,k+1/2}$, $\tilde{E}^{n+1/2}_{y,i+1/2,j,k+1/2}$ and $\tilde{E}^{n+1/2}_{z,i+1/2,j+1/2,k}$ 
constructed by our MEC strategy, 
the final CT update evolves the cell face-centered magnetic fields satisfying the $\nabla
\cdot \mathbf{B} = 0$ condition on a staggered grid. Displaying only in $x$-direction, we have
\begin{eqnarray}
{b}^{n+1}_{x,i+1/2,j,k}&&=
{b}^{n}_{x,i+1/2,j,k}\nonumber\\
&&-\frac{\Delta t}{\Delta y}
\Bigr\{
 \tilde{E}^{n+1/2}_{z,i+1/2,j+1/2,k} 
-\tilde{E}^{n+1/2}_{z,i+1/2,j-1/2,k}
\Bigl\} 
-\frac{\Delta t}{\Delta z}
\Bigr\{
-\tilde{E}^{n+1/2}_{y,i+1/2,j,k+1/2} 
+\tilde{E}^{n+1/2}_{y,i+1/2,j,k-1/2}
\Bigl\}.
\label{sec2:InductionBx2D-final}
\end{eqnarray}
This completes our description of all procedures in the 3D USM algorithm for a single time step update. 

\section{Summary}
\label{sec:summary}
We summarize the 3D USM algorithm as follows:

\begin{enumerate}
\item Calculate the normal predictor states in all $x,y,z$-directions using
the algorithm described in Section \ref{sec:normal-predictor-dataReconst}.
When calculating the normal state in each direction, include the associated
MHD term that is proportional to the gradient of the normal field, 
see the first relation in (\ref{sec2:NormalPredictorReduced}). During
each normal predictor calculation, the eigensystems
in the normal direction are to be computed. They are
stored for later use in the transverse correctors. At the same time, the
summations of the jumps in all characteristic variables 
are also computed and stored, see Equations
(\ref{Eqn-section2:TransFluxGradients})-(\ref{Eqn-section2:upwindSlopeLimiter}),
the pseudo-code in Section \ref{sec:CharacteristicTracingForTransverse}, the
sigma summations in Equations
(\ref{sec2:Transverse-y-EW-approx-3dcorrect_c}) and
(\ref{sec2:Transverse-z-EW-approx-3dcorrect_b}).

\item The normal predictor states are updated via the transverse correctors
described in Section \ref{sec:reduced3dCTU-usm}. This step uses two of the
stored sigma summation terms that are calculated and stored in each normal predictor
step. The summations reflect the transverse flux gradients using the
characteristic tracing approach. 
\begin{enumerate}
\item The reduced 3D CTU scheme then proceeds to advance the normal
fields by half a time step using CT as illustrated in Section
\ref{sec:CThalfTimeStepUpdate}, finalizing all the interface state
calculations. In the reduced 3D CTU scheme, the formal stability limit is
given by a CFL number that is less than $\frac{1}{2}$. 
\item If the full 3D CTU scheme is chosen, the algorithm needs to take one
more correction step, presented in Section \ref{sec:full3dCTU-usm}.
This correction step is essential in order to provide the full stability
limit by including the diagonally moving upwind
information along the corners in a 3D control volume. 
Similar to the reduced CTU scheme, the full 3D
CTU scheme is completed by evolving the cell face-centered magnetic
fields by half a time step as in Section \ref{sec:CThalfTimeStepUpdate}. 
\end{enumerate}
So far, both of the two CTU algorithms have
required the first set of three Riemann problems that are used to advance the
magnetic fields by CT.

\item Solve the final set of three Riemann problems at cell interfaces and
update the cell-centered conservative variables to the next time step as
described in Section \ref{sec:unkUpdate}. The total number of Riemann
solves therefore becomes six\footnote{Our unsplit data
reconstruction-evolution algorithm can be easily modified for use as a gas
hydrodynamics solver by omitting those steps related to the magnetic fields. In
this case, there are only three Riemann solves required as there is no CT
update needed. This unsplit hydrodynamics solver has been also available in
FLASH's official releases.}.

\item Calculate the electric fields at cell corners by using the upwind-MEC
algorithm described in Section \ref{sec:upwindMEC}. With these electric fields,
the magnetic fields at cell face centers are updated to the next time
step by CT. The cell-centered magnetic fields are updated by taking
arithmetic averages of these divergence-free magnetic fields at cell face
centers (e.g., Equation (\ref{sec2:BxCenterReconstruction}) in Section
\ref{sec:normal-predictor-dataReconst}).

\end{enumerate} 

\section{Numerical Results}
\label{sec:results}
In this section we exhibit the
accuracy, stability, convergence and
computational performance of the USM scheme on a suite of 3D MHD problems.
These results
show that the scheme is very robust with the full CFL
stability bound. The scheme is second-order accurate for smooth flows and
maintains the solenoidal constraint on the magnetic field up to machine
round-off error.  The full 3D CTU method is our primary default method for which 
a CFL number of 0.95 is chosen in all of the simulations presented here. 
We also show a set of comparison studies
between the reduced and full 3D CTU schemes.
Insofar as choice of Riemann solver is concerned, we use the Roe-type linearized solver
[\citenum{Roe1981,Toro2009}] and the HLLD solver
[\citenum{MiyoshiKusano2005}]. Our choices 
for the normal predictor step are MUSCL-Hancock, PPM, and WENO5.

\subsection{Field Loop Advection}
\label{sec4:fieldLoop}
This problem is notoriously difficult to solve, not because of any strong shock causing numerical
instability and leading to code to crash, but rather because it requires full accounting of multidimensional 
advection in a stable matter, such as including the multidimensional MHD 
terms [\citenum{GardinerStone2005,GardinerStone2008,LeeDeane2009,MiniatiMartin2011}].
Failure to do so results in an erroneous 
generation of in-plane magnetic field, 
which results in the distortion of the initially
circular (2D) or cylindrical (3D) field loop.

In addition to the standard field loop advection case studied in [\citenum{GardinerStone2008}],
we also consider a small angle advection case. This turns out to be a much more stringent test than the
standard advection configuration which assumes a (relatively) large angle between the advection flow and the Euclidean coordinate axes.
In the small angle configuration there is one dominating coordinate direction along 
which the field loop is advected.
This means that, in practice, there is only one direction from which a numerical scheme can obtain the numerical dissipation
required for stability. In multidimensional problems, inadequate numerical dissipation from transverse directions can give rise to anomalous oscillatory behavior in physical variables.

We begin by describing the initial setup for the standard large angle advection case 
following the configuration of [\citenum{GardinerStone2008}].
The weakly magnetized 3D cylindrical field loop is initialized with a very high plasma beta $\beta=p/B_p=2\times10^6$ in the inner region, where $B_p=(B_1^2+B_2^2+B_3^2)/2$. Inside the loop the magnetic field strength is very weak and the flow dynamics are dominated by the gas pressure.

The initial field loop is  tilted around  the $x_2$ (or $y$) axis by $\omega =\tan^{-1}\Omega$ radians in a 3D periodic box $[-0.5,0.5]\times[-0.5,0.5]\times[-1,1]$. For the standard large angle setup, we choose $\Omega=2$.
The field loop is frozen into the ambient plasma and is advected diagonally across the domain with the plasma
advection velocity $(u,v,w)=(1,1,2)$. The density and pressure are equal to unity everywhere, and $\gamma=\frac{5}{3}$. 

The magnetic field components are initialized by taking numerical curl
of the magnetic vector potential $\mathbf{A}=(A_1,A_2,A_3)^T$ in order to ensure
$\nabla \cdot \mathbf{B} = 0$ initially. 
The relationship between magnetic field and vector potential gives
\begin{equation}
B_1= \frac{\partial A_3}{\partial x_2} - \frac{\partial A_2}{\partial x_3}, \;\;\;\; 
B_2=-\frac{\partial A_3}{\partial x_1} +\frac{\partial A_1}{\partial x_3}, \;\;\;\;
B_3= \frac{\partial A_2}{\partial x_1} - \frac{\partial A_1}{\partial x_2}.
\end{equation}
For the components of $\mathbf{A}$ we choose $A_1 = A_2 = 0$ and
\begin{eqnarray}
A_3=
\cases{A_0\left(R-r\right) \;\;\;\; \mbox{if} \;\;\; r\leq R,\cr
       0\;\;\;\;\;\;\;\;\;\;\;\;\;\;\;\;\;\;\;\; \mbox{otherwise.}}
\label{Eqn:VectorPotentialAz}
\end{eqnarray}
By using this initialization process divergence-free magnetic fields are
well constructed numerically on a staggered grid. The parameters in
Equation (\ref{Eqn:VectorPotentialAz}) are $A_0=10^{-3}$ and $R=0.3$. 

The two coordinate systems $(x_1,x_2,x_3)$ and $(x,y,z)$ are related by a rotation about the $y$-axis,
which is given by

\begin{eqnarray}
\left( \begin{array}{ccc}
x_1 \\
x_2 \\
x_3
\end{array} \right)
=
\left( \begin{array}{ccc}
\mbox{cos}\omega & 0 & \mbox{sin}\omega \\
0 & 1 & 0 \\
-\mbox{sin}\omega & 0 & \mbox{cos}\omega 
\end{array} \right)
\left( \begin{array}{ccc}
x \\
y \\
z
\end{array} \right).
\label{Eqn:FL_RotationMatrix}
\end{eqnarray}

Before we present the standard large angle field loop advection, we first consider
a small angle advection case in 2D. The 2D initial conditions can be found in 
[\citenum{GardinerStone2005,LeeDeane2009}] and we will not repeat the details here.
In the 2D setup, the velocity field is given by
\begin{equation}
\label{Eqn:2DFieldLoopInitialVelocity}
\mathbf U=(u_0\mbox{cos}\theta,u_0\mbox{sin}\theta,1)^T,
\end{equation}
where $u_0=\sqrt{5}$. We chose $\theta=\mbox{tan}^{-1}(0.01)\approx0.573^{\circ}$ for a small angle advection test.
With this setup, the field loop is advected almost entirely in the positive $x$-direction.
This situation makes it hard to stabilize the solution during advection because there is not enough numerical
dissipation from the $y$-direction. The solution behavior completely relies on the dissipation mechanism from
the $x$-direction only, making the problem very unstable if no special care is taken to stabilize it.

\begin{figure}[htbp]
  \centerline{
    {\subfigure[$B_p$  at $t=0.1$]{\includegraphics[width=3.5in]{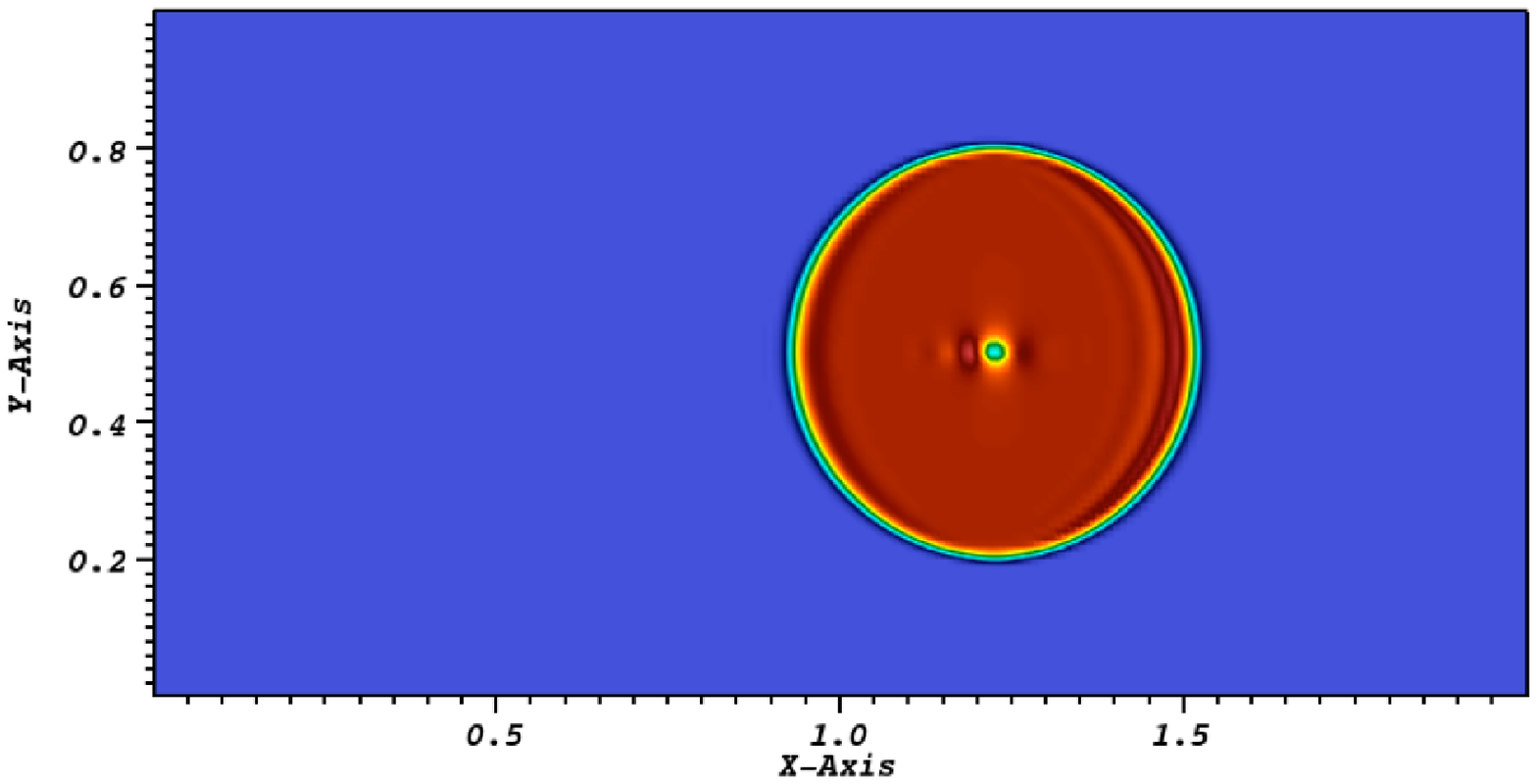}}}
    {\subfigure[$B_p$  at $t=2$]{\includegraphics[width=3.5in]{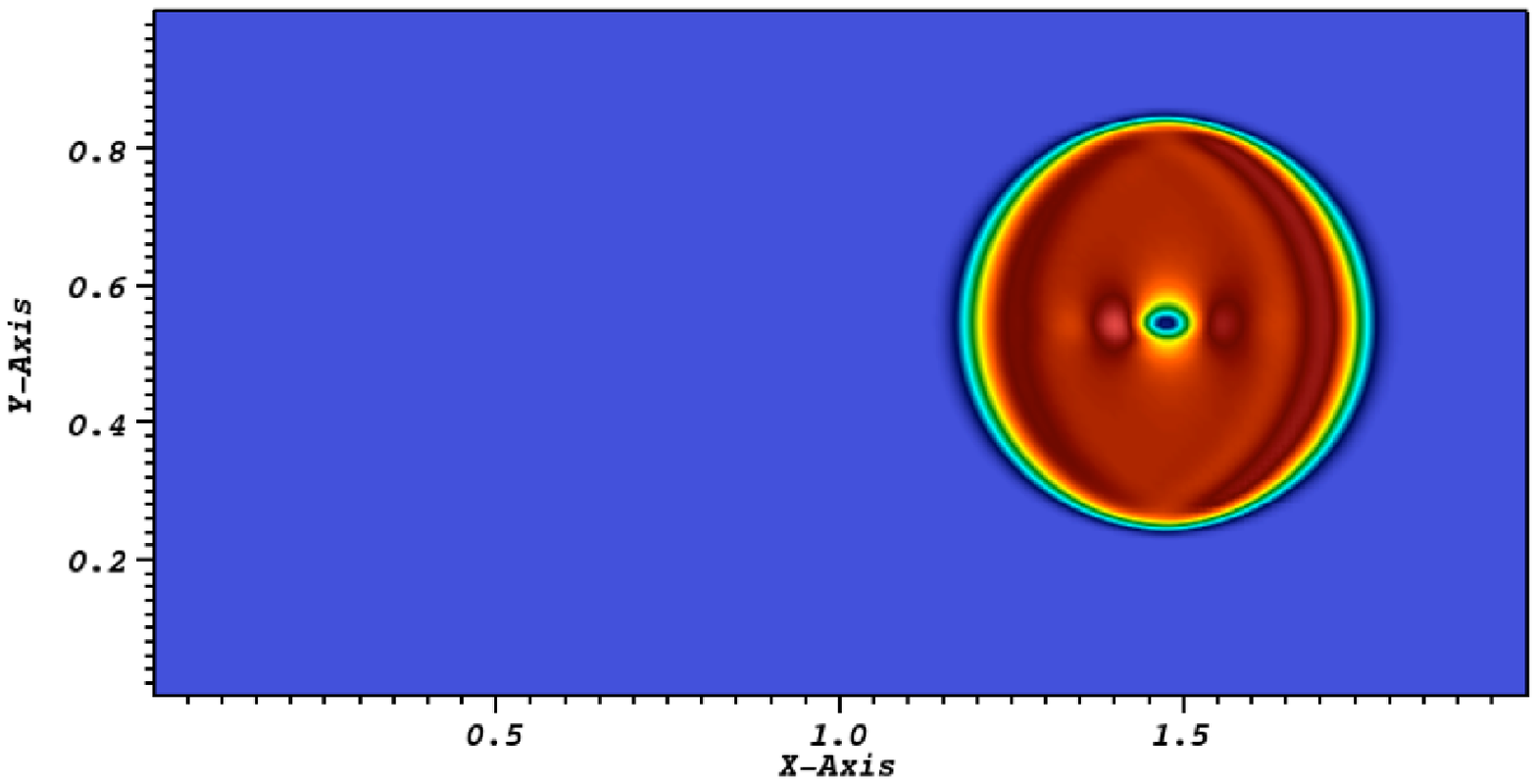}}}
    }
\caption{The 2D field loop advection using a small advection angle $\theta\approx0.573^{\circ}$ relative to the $x$-axis. The images are magnetic pressures at times $t=0.1$ and $2$ using PPM and the Roe Riemann solver. The minmod slope limiter is used for taking slope gradients of characteristic variable in the PPM reconstruction step. All results are resolved on $200\times100$ grid cells using the upwind-MEC scheme.}
\label{FieldLoopAdvect2DSmallAngle}
\end{figure}

In Figure \ref{FieldLoopAdvect2DSmallAngle}, we illustrate the evolution of the magnetic pressure $B_p$ at $t=0.1$ and $2.0$ using 
the upwind-MEC scheme described in Section \ref{sec:upwindMEC}.
In consequence of the upwinding dissipation mechanism, the upwind-MEC scheme stabilizes the solutions 
extremely well, suppressing the anomalous behavior during the advection.

\begin{figure}[htbp]
  \centerline{   
    {\subfigure[$B_p$ at  $t=2$.]
        {\epsfig{file=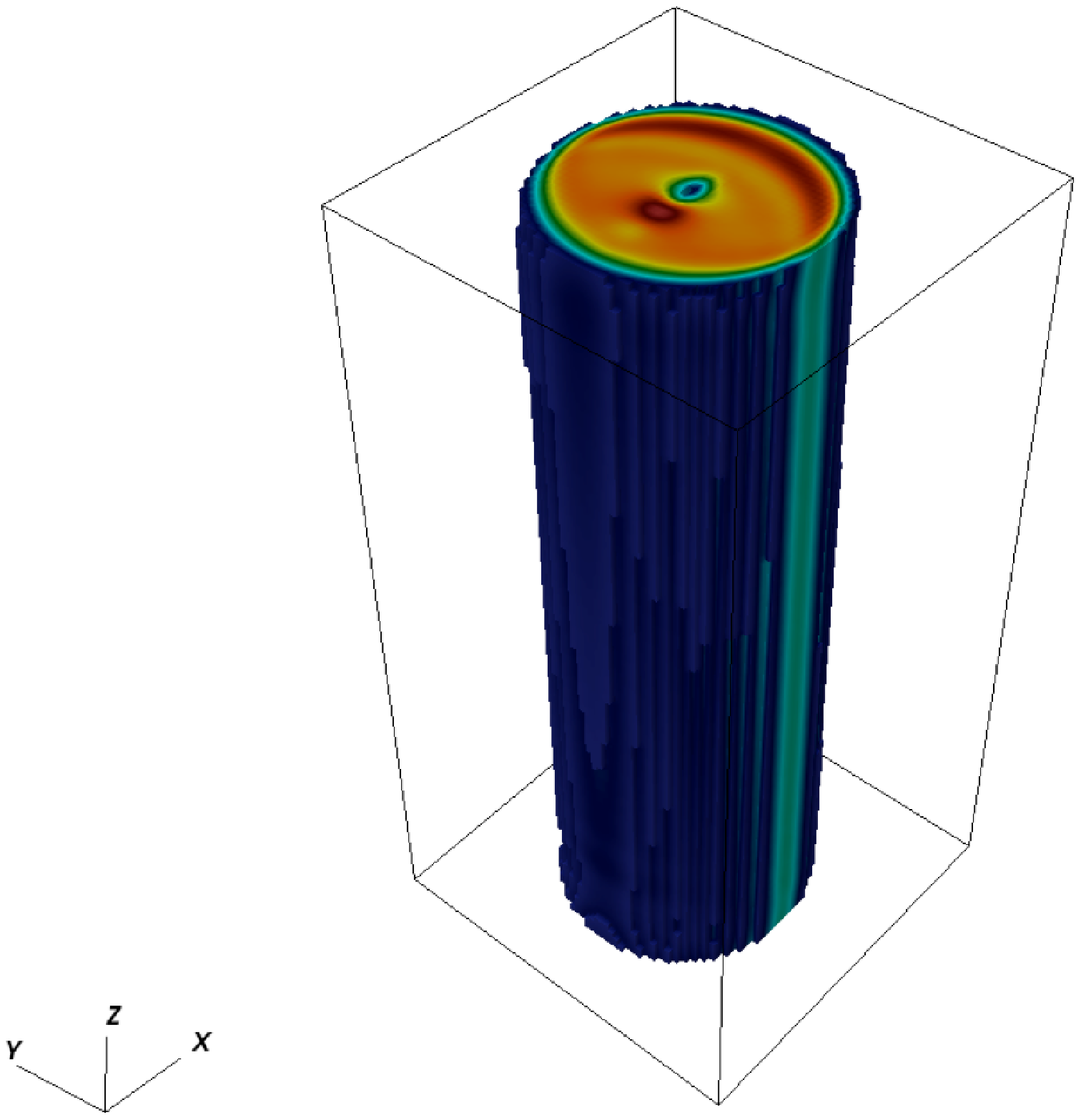, width=2.2 in}}}
    {\subfigure[$B_p$ at  $t=2$.]
        {\epsfig{file=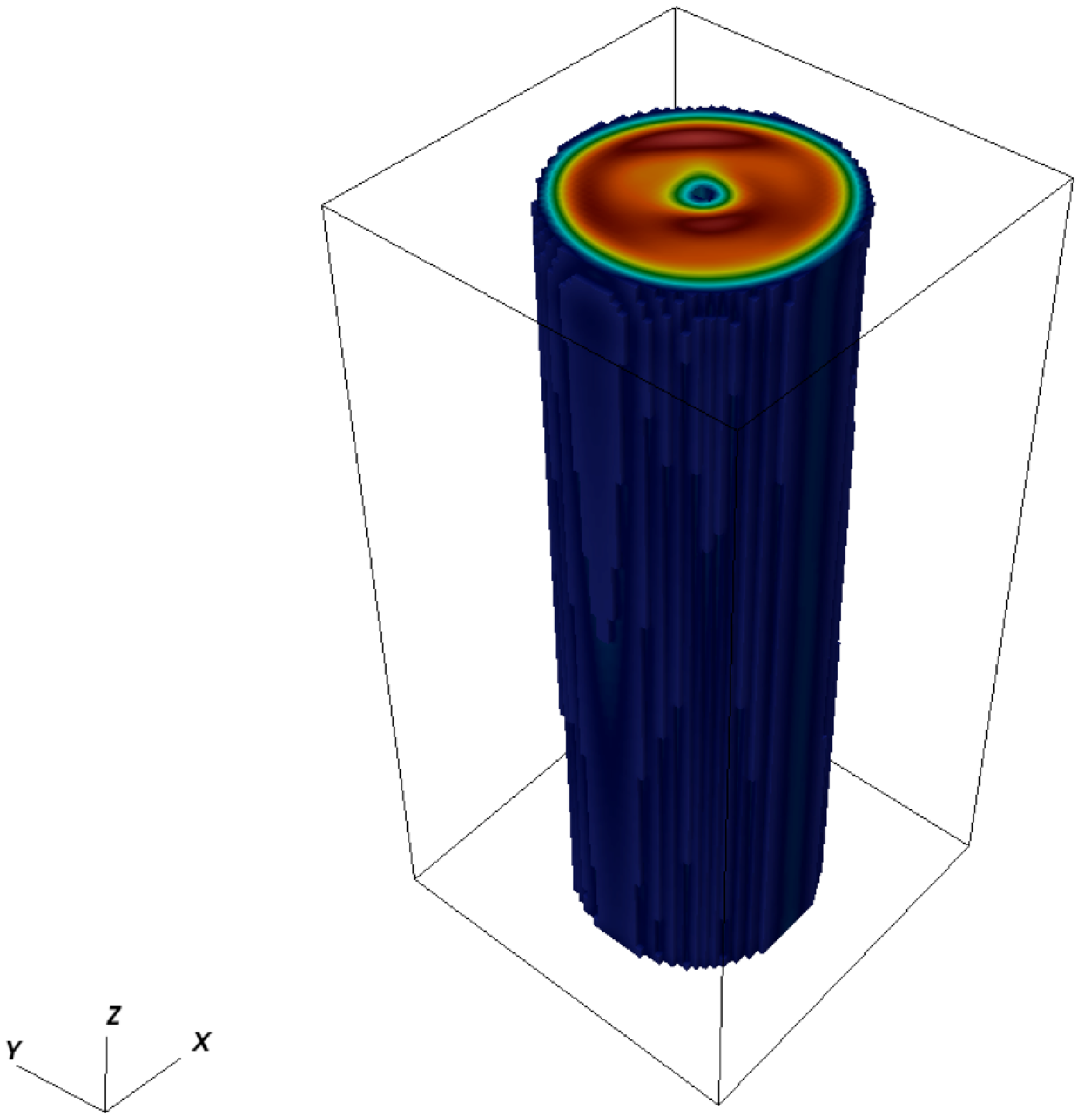,  width=2.2 in}}}
        {\subfigure[$B_p$ at $t=1$.]
        {\epsfig{file=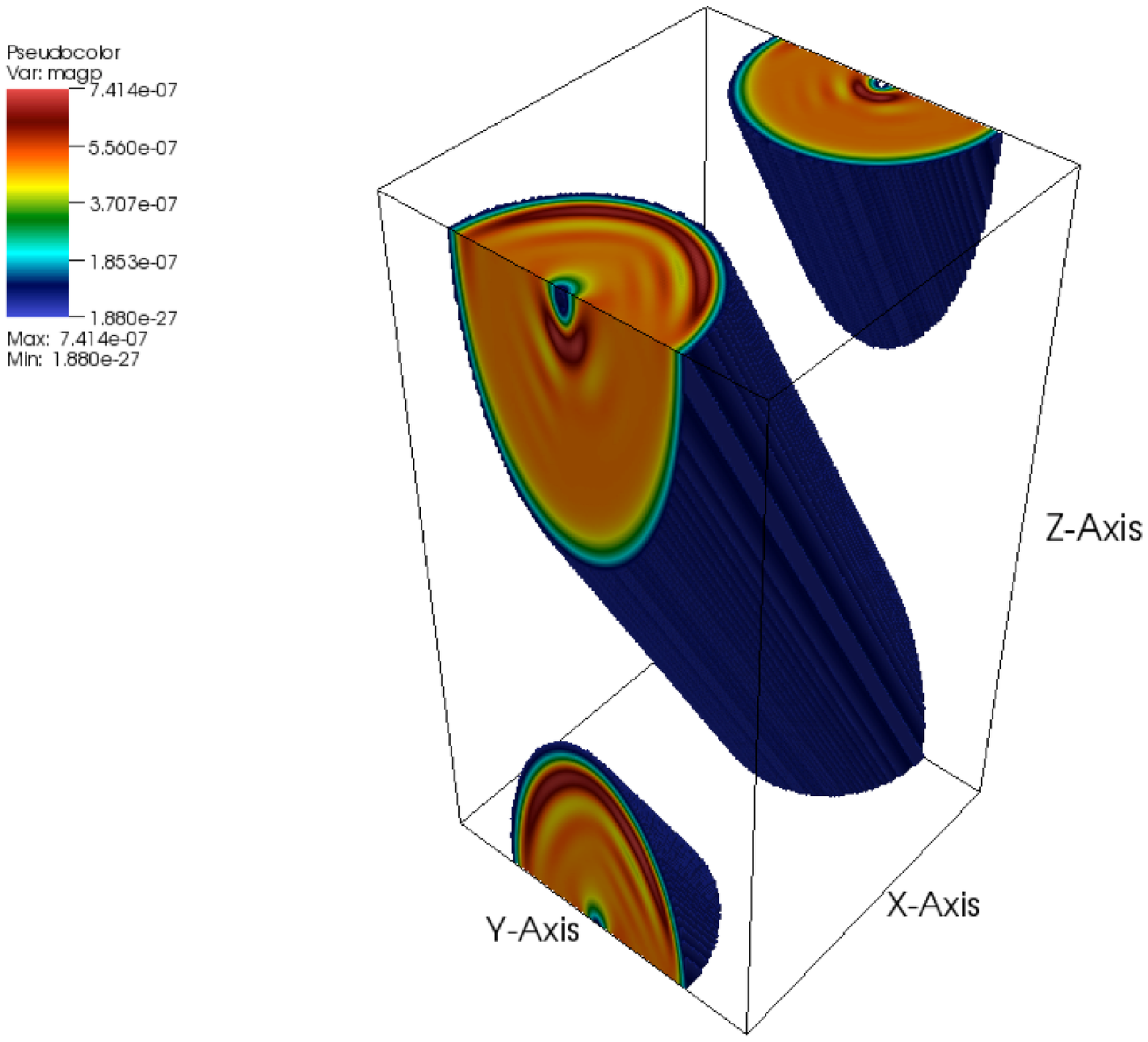, height=2.2 in, width=2.4 in}}}
    }
    \caption{(a) 3D field loop advection using a small advection angle $\theta\approx0.573^{\circ}$ relative to the $x$-axis. 
    (b) 3D field loop advection using a large advection angle using $\mathbf U=(1,1,2)^T$. 
    (c) The standard field loop advection problem at time $t=1$.
    All results use PPM and the Roe Riemann solver, and the minmod slope limiter on characteristic variables in the PPM reconstruction. 
    The results in (a) and (b) are resolved on $64\times64\times128$  grid cells, and (c) on $128\times128\times256$. The upwind-MEC is used in all cases.}
\label{FieldLoopAdvect3D}
\end{figure}

In Figure \ref{FieldLoopAdvect3D} (a), the small angle advection test is repeated in 3D,
whereas a large angle advection is demonstrated in Figure \ref{FieldLoopAdvect3D} (b).
The velocity fields are respectively given by 
$\mathbf U=(\mbox{cos}\theta,\mbox{sin}\theta,2)^T$ and $\mathbf U=(1,1,2)^T$ for the small and large angle runs.
In (a), the same small advection angle $\theta\approx0.573^{\circ}$ was used relative to the $x$-axis as in the 2D case.
In the large angle case in (b), the field loop makes a domain diagonal advection from the given initial velocity condition.
We set the tilt angle $\omega$ in Equation (\ref{Eqn:FL_RotationMatrix})  to be same as $\theta$ for both (a) and (b).
In both runs, we show that the upwind-MEC scheme manifests an oscillation-free advections, well-preserving 
the initial cylindrical shape in the magnetic pressure as shown.
As noted above, 
numerical dissipation in the large angle run is naturally added from all directions, rather than from
one specific direction along the advection in the small angle case. 
Such added dissipation makes the large angle case easier to demonstrate than the small angle case.

As a final test in this section, we perform the standard field loop advection problem in Figure \ref{FieldLoopAdvect3D} (c),
following the configuration in [\citenum{GardinerStone2008}]. 
We see that the upwind-MEC scheme performs very well in evolving the field loop successfully to the final time $t=1$.
This result in (c) can be directly compared to the results reported in [\citenum{GardinerStone2008}].
We also report that the upwind-MEC scheme increases
the maximum value of the magnetic pressure by  $48\%$ to $7.41\times10^{-7}$ from its initial value of $5\times10^{-7}$. 
The larger growth of the maximum value is found in the standard-MEC scheme, increasing the initial value by $69\%$ (not shown here).  

\begin{figure}[htbp]
  \centerline{   
    {\subfigure[Normalized, volume averaged magnetic energy density in time]
        {\epsfig{file=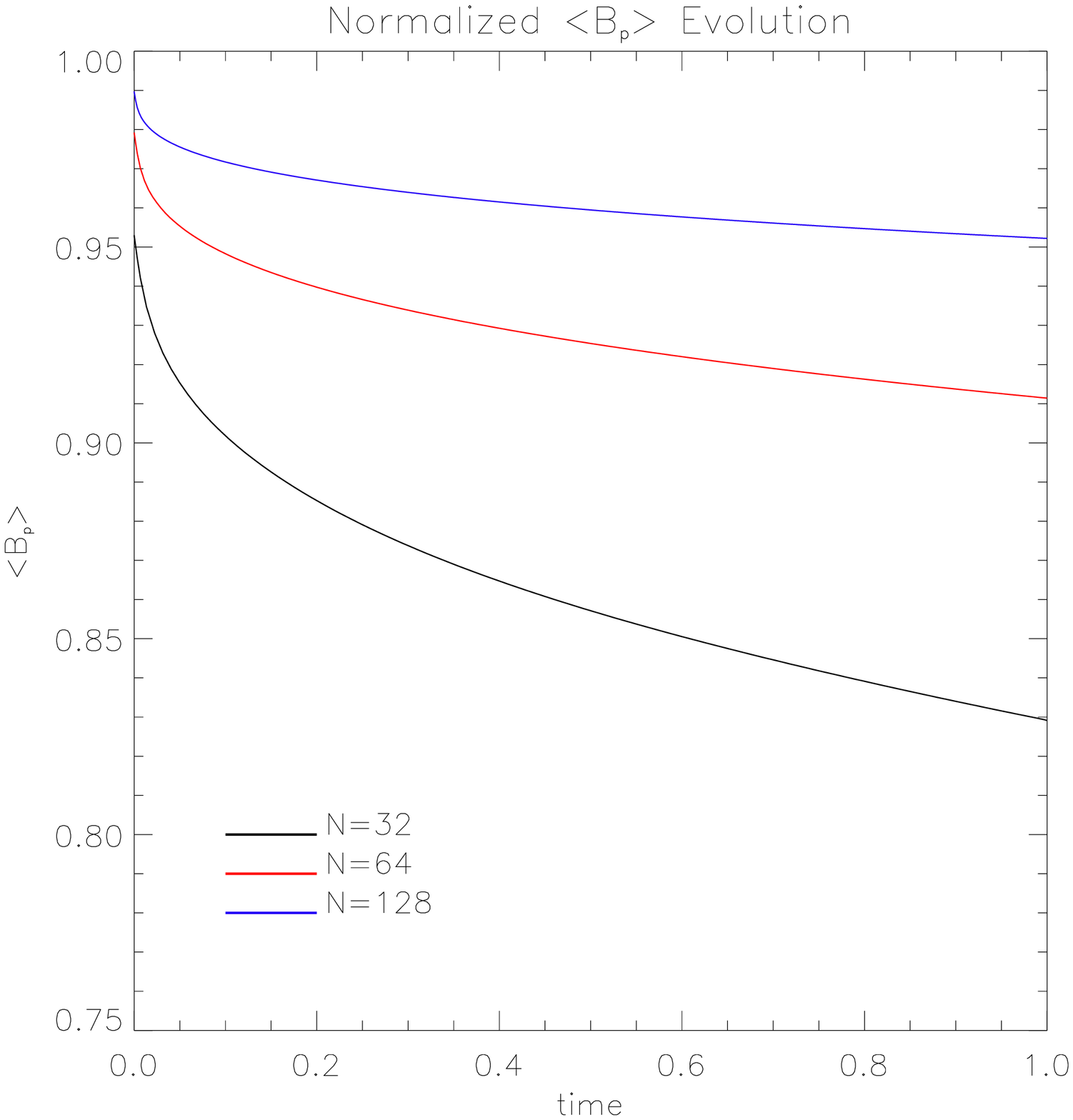, height=3.5 in, width=3.5 in}}}
    {\subfigure[Normalized $B_3$ error in time]
        {\epsfig{file=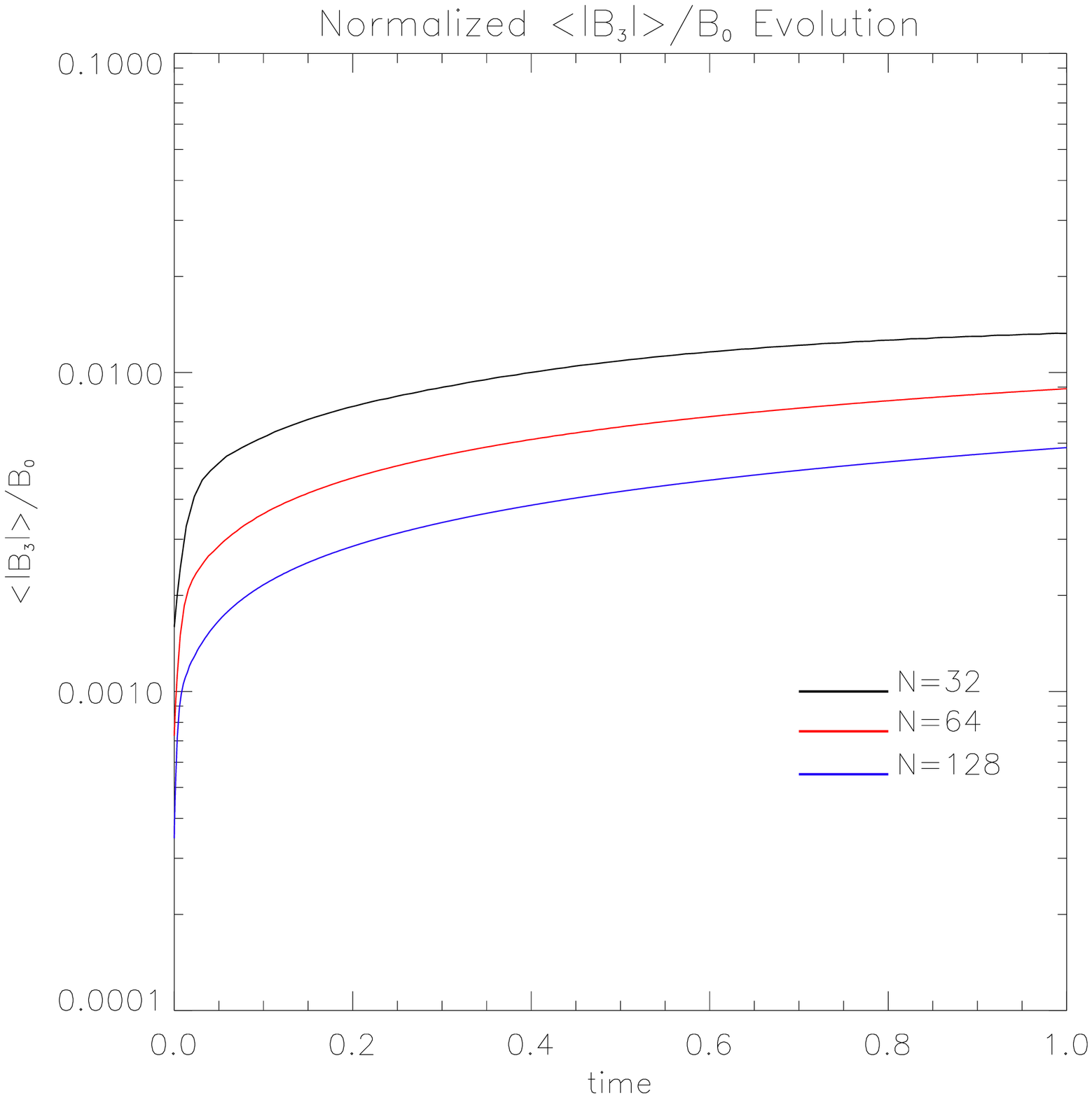, height=3.5 in, width=3.5 in}}}
    }
    \caption{Time evolution of (a) the normalized, volume averaged magnetic energy density $<B_p>=<B^2>$ and (b) the 
   normalized error $<|B_3|>/B_0$. Three different results on the grid resolutions of $N=32, 64$ and $128$ are plotted.
   The full CTU scheme is adopted with CFL=0.95 using PPM and the Roe Riemann solver.}
\label{FieldLoopAdvectQuantitative}
\end{figure}

Furthermore, we present two quantitative results in Figure \ref{FieldLoopAdvectQuantitative}. They include (a) the temporal evolution of the 
volume-averaged magnetic energy density normalized to the initial (analytic) value $<B_p>=<B^2>=B_0^2\sqrt{5}\pi R^2/2$; and (b) the temporal evolution of the normalized error $<|B_3|>/B_0$. Both results in (a) and (b) are similar to those reported in [\citenum{GardinerStone2008,MiniatiMartin2011}]. However in (b), the final values at $t=1$ are found out to be little larger than those in [\citenum{GardinerStone2008,MiniatiMartin2011}]  at each grid resolution. This is probably because our full 3D CTU method of including the MHD multidimensional terms ignores the $\mathcal{O}(\Delta t)$ terms in evaluating the eigensystems of the $\mathbf{A}_d$ matrices at $n+\frac{1}{3}$ (see for example Equation (\ref{eqn:first-order-matrixEvaluation})), where $d=x,y,z$.

\subsection{Circularly Polarized Alfv\'{e}n Wave}
\label{sec4:alfvenWave}
In the next test  we solve the circularly polarized Alfv\'{e}n wave and its
propagation [{\citenum{Toth2000,GardinerStone2005,GardinerStone2008}].
This problem provides an important quantitative test of the 3D USM scheme
because the smooth initial conditions are nonlinear solutions to the problem.
The Alfv\'{e}n wave propagates parallel to the
$x_1$-axis of a transformed coordinate system $(x_1,x_2,x_3)$ in
the periodic computational domain $[0,3]\times[0,1.5]\times[0,1.5]$. 
The computational domain is
resolved on $2N\times N \times N$ grid cells, where we adopt $N=8,16,32$ and $64$
for the convergence study.

The relationship between the rotated coordinated system $(x,y,z)$ and the non-rotated system
$(x_1,x_2,x_3)$ is 
described by the following coordinate transformation

\begin{eqnarray}
\left( \begin{array}{ccc}
x_1 \\
x_2 \\
x_3
\end{array} \right)
=
\left( \begin{array}{ccc}
 x\cos\alpha\cos\beta +y \cos\alpha\sin\beta + z\sin\alpha\\
-x\sin\beta +y \cos\beta\\
-x\sin\alpha \cos\beta -y \sin\alpha\sin\beta + z\cos\alpha\\
\end{array} \right),
\label{Eqn:CPAW_RotationMatrix}
\end{eqnarray}
where $\sin\alpha=\frac{2}{3}$, $\sin\beta=\frac{2}{\sqrt{5}}$,  $\cos\alpha=\frac{\sqrt{5}}{3}$, 
and $\cos\beta=\frac{1}{\sqrt{5}}$.

The initial conditions we use are the same as the equivalent test problems
described in [\citenum{GardinerStone2008}]. The initial magnetic field is given by
\begin{equation}
\mathbf{B}=\left(B_{x_1},B_{x_2},B_{x_3}\right)^T=\bigl(1,0.1\sin(2\pi x_1/\lambda),0.1\cos(2\pi x_1/\lambda)\bigr)^T,
\label{alfven_B}
\end{equation}
and similarly the velocity field is
\begin{eqnarray}
\mathbf{U}=\left(U_{x_1},U_{x_2},U_{x_3}\right)^T=
\cases{\bigl(0,0.1\sin(2\pi x_1/\lambda),0.1\cos(2\pi x_1/\lambda) \bigr)^T \;\;\;\; \mbox{for traveling wave}, \cr
             \bigl(1,0.1\sin(2\pi x_1/\lambda),0.1\cos(2\pi x_1/\lambda) \bigr)^T \;\;\;\; \mbox{for standing wave.}} 
\end{eqnarray}
We set the wavelength $\lambda=1$. The density and the gas pressure are initialized by $\rho=1$ and $p=0.1$.
We choose PPM and the HLLD Riemann solver, with the monotonized central (MC) limiter.

Figure \ref{fig:cpaw_conv} (a) and (b) show the numerical errors on a logarithmic
scale obtained with four different grid resolutions of $N=8,16,32$ and $64$.
We test the reduced 3D CTU scheme using CFL=0.475 
and
the full 3D CTU scheme using CFL=0.475 and 0.95 
for the convergence study.
The errors of the standing and traveling waves are plotted in (a) and (b) respectively. 
For all cases we follow the error calculation formula used by 
Gardiner and Stone [\citenum{GardinerStone2008}] in order to compare our results
with theirs. 
The results in Figure \ref{fig:cpaw_conv} (a) and (b) show a second-order convergence rate of both
reduced and full 3D CTU schemes for the smooth Alfv\'{e}n wave problem.

We also measure the relative 
CPU cost of the full CTU scheme to the reduced CTU scheme, 
$\mbox{CPU}_{\mbox{rel}}=\mbox{CPU}_{\mbox{f-ctu}}/\mbox{CPU}_{\mbox{r-ctu}}$.
We find that $\mbox{CPU}_{\mbox{rel}}$ is about 0.8 on average, which indicates that our full CTU scheme 
with a higher CFL number (e.g., 0.95) is $20\%$ more computationally efficient than 
the reduced CTU with a lower CFL number (e.g., 0.475).
The equivalent performance comparison is different in the 6-solve and the 12-solve algorithms in 
[\citenum{GardinerStone2008}] in that their relative CTU performance turns out to be 1.

As the error magnitudes are nearly identical for both
standing and traveling wave modes in the reduced CTU with CFL=0.475
and the full CTU with CFL=0.95, we conclude that our full 3D CTU scheme exhibits
better performance while providing numerical solutions that are second-order accurate.

The figures exhibit a dependence of the truncation error in the full CTU scheme on CFL, in both wave modes. 
The error corresponding to CFL=0.475 is smaller in the standing wave, whereas it is larger
in the traveling wave simulation. 
This type of CFL dependence is also seen in [\citenum{GardinerStone2008}].

\begin{figure}[htbp]
  \centerline{   
    {\subfigure[Convergence rate for the standing wave solutions at $t=1.0$]
        {\epsfig{file=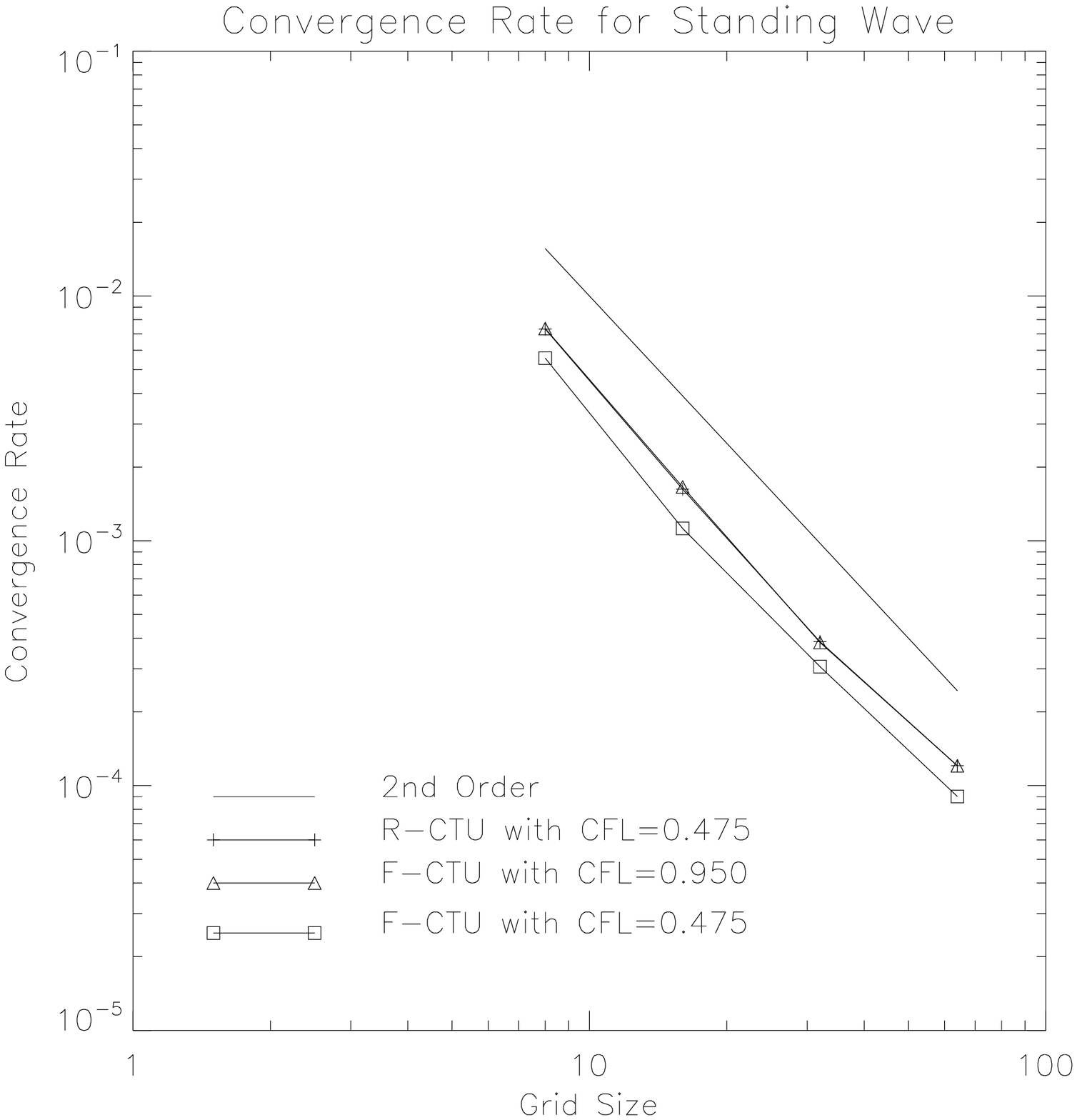, height=3.3 in, width=3.5 in}}}
    {\subfigure[Convergence rate for the traveling wave solutions at $t=1.0$]
        {\epsfig{file=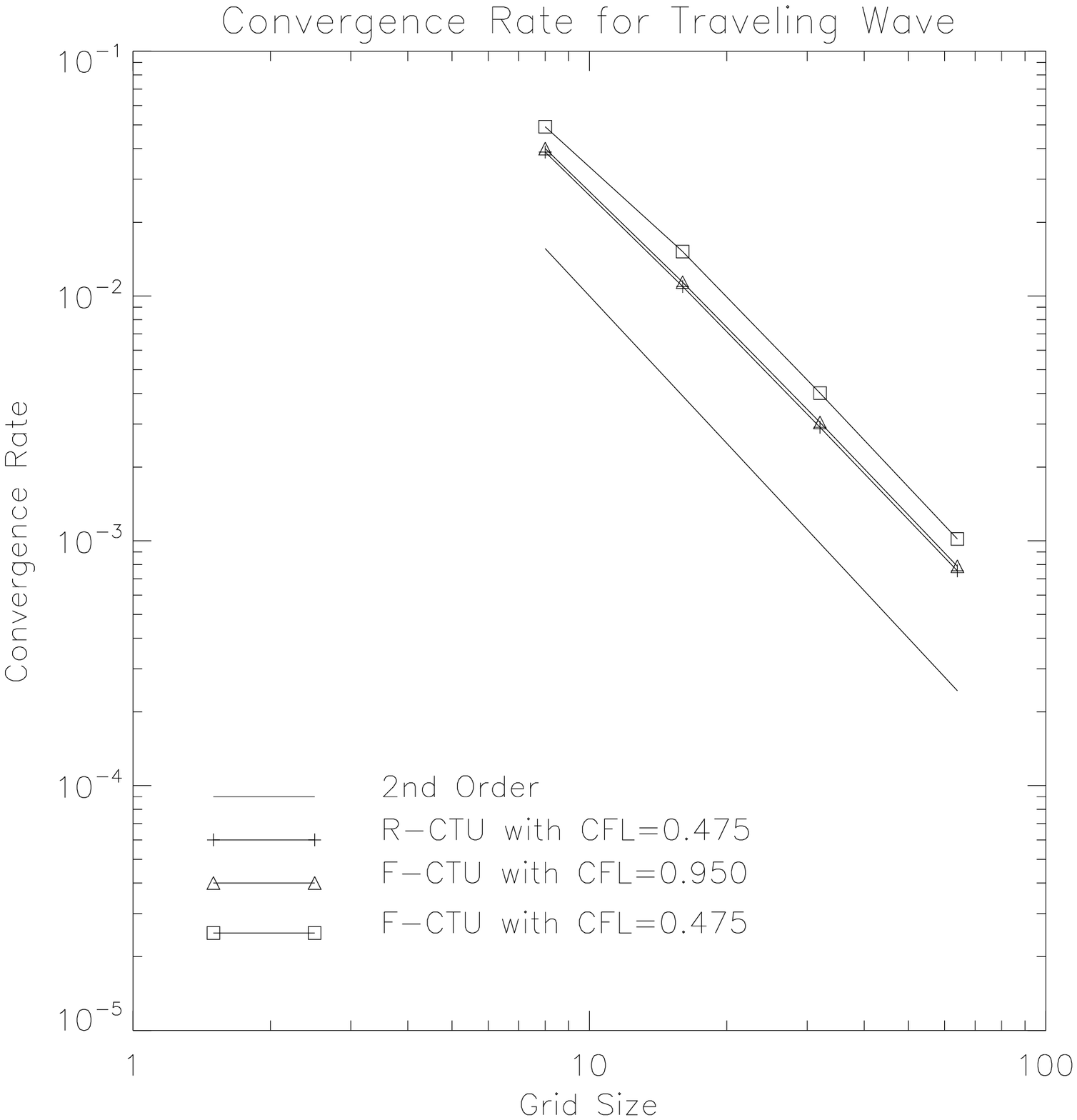, height=3.3 in, width=3.5 in}}}
    }
    \caption{The circularly polarized Alfv\'{e}n wave convergence rate for both
the standing and traveling wave problems. PPM is used along with the HLLD Riemann solver.}
\label{fig:cpaw_conv}
\end{figure}

\subsection{Orszag-Tang Problem}
\label{sec4:orszagTang}
The third test problem is the Orszag-Tang MHD vortex problem
[\citenum{OrszagTang1979}]. We follow the 3D extension [{\citenum{Helzel2011}}] of the
2D problem in which the initial velocity field is slightly perturbed by $\epsilon$ in the vertical direction.
That is, the initial velocity field defined on a periodical computational domain $[0,1]\times[0,1]\times[0,1]$ is written as
\begin{equation}
\mathbf U=(-(1+\epsilon \sin 2\pi z)\sin 2\pi y,(1+\epsilon \sin 2\pi z)\sin 2 \pi x, \epsilon \sin 2\pi z)^T,
\end{equation}
where we use $\epsilon=0.2$ as in  [{\citenum{Helzel2011}}].
The rest are initialized similar to the 2D case so that 
\begin{equation}
\rho = \gamma ^ 2,\; p=\gamma,\; \mathbf B=(-\sin 2\pi y,\sin 4 \pi x,0)^T,
\end{equation}
where $\gamma=\frac{5}{3}$.
As in the 2D case, the plots in Figure \ref{OrszagTangDens} exhibit
nonlinear steepening that builds strong discontinuities from the smooth initial
conditions. We show the evolutions of density at $t=0.5$ and $1.0$
on $128^3$ grid cells. The density images at the top of the domain
are very similar to those in the standard 2D case at each corresponding time 
(e.g., see [\citenum{LeeDeane2009}]). 
\noter{The flow symmetries are also well preserved
in (b) where density has developed into more complicated discontinuous flows.}
The Roe Riemann solver is used with the PPM scheme
for data reconstruction-evolution in normal direction with MC limiter.

\begin{figure}[htbp]
  \centerline{   
    {\subfigure[Density at $t=0.5$]{\epsfig{file=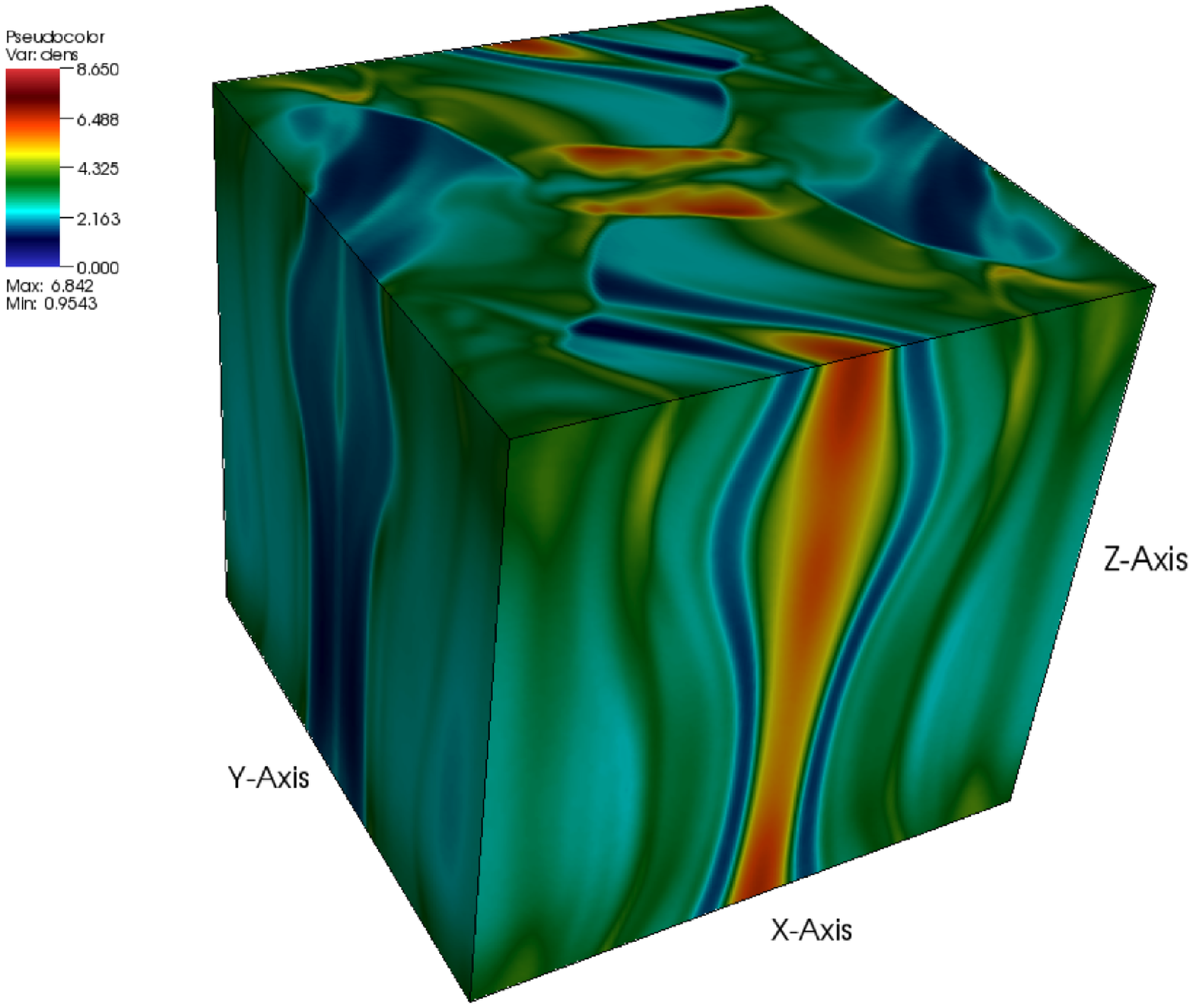, height=3.0 in, width=3.5 in}}}
    {\subfigure[Density at $t=1.0$]{\epsfig{file=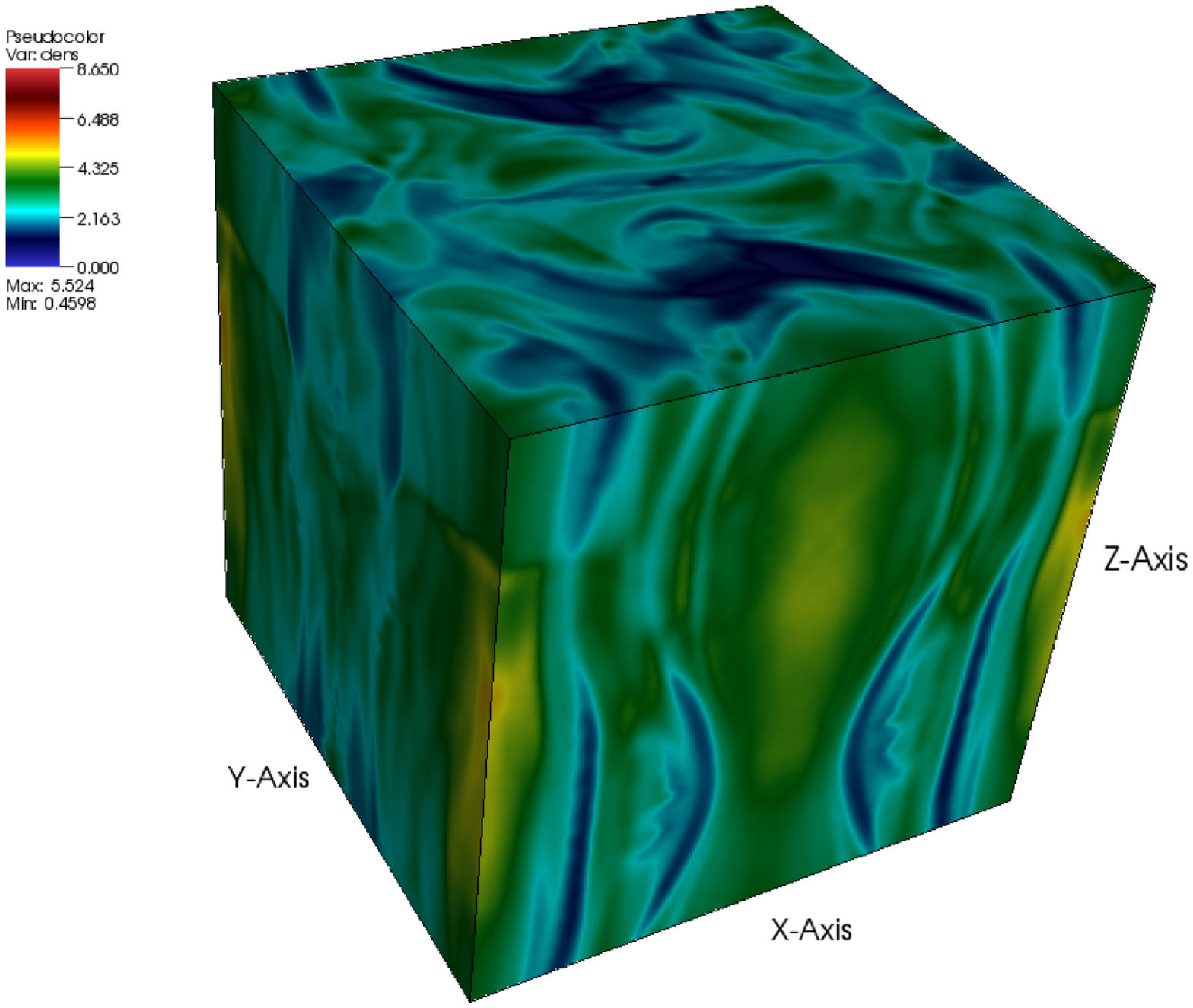, height=3.0 in, width=3.5 in}}}
    }
\caption{Density  plots of the Orszag-Tang problem at a resolution of $128^3$.}
\label{OrszagTangDens}
\end{figure}

\subsection{Rotor  Problem}
\label{sec4:rotor}
We extend the 2D rotor problem [\citenum{BalsaraSpicer1999,Toth2000,LeeDeane2009}] to a 3D case by applying a small velocity perturbation analogous to that introduced in the 3D Orszag-Tang problem in Section \ref{sec4:orszagTang}.
A dense rotating cylinder is initialized on a unit cube domain $[0,1]\times[0,1]\times[0,1]$ with non-reflecting boundary conditions. The initial velocity field is defined by
\begin{equation}
\mathbf U=(u_{2d}(1+\epsilon \sin 2\pi z),v_{2d}(1+\epsilon \sin 2\pi z), \epsilon \sin 2\pi z)^T,
\end{equation}
where $\epsilon=0.3$ and 
\begin{eqnarray}
u_{2d}   &=&\cases{-f(r)u_0(y-0.5)/r_0 & for   $r \leq r_0$\cr
		       -f(r)u_0(y-0.5)/r      & for   $r_0 < r < r_1$\cr
		        0                               & for   $r \geq r_1$\cr},\\ \nonumber \\
v _{2d}  &=&\cases{ f(r)u_0(x-0.5)/r_0 & for   $r \leq r_0$\cr
		        f(r)u_0(x-0.5)/r      & for   $r_0 < r < r_1$\cr
		        0                              & for   $r \geq r_1$\cr}.\\ \nonumber
\end{eqnarray}
The density, pressure, magnetic field, and the parameters are initialized as in the standard 2D case given by
\begin{equation}
\rho=\cases{10                  & for   $r \leq r_0$\cr
		         1+9f(r)           & for   $r_0 < r < r_1$\cr
		         1                    & for   $r \geq r_1$\cr},
\end{equation}
\begin{equation}
p  =1, \; \mathbf B=(5/\sqrt{4\pi},0,0)^T,
\end{equation}
where $u_0=2, r_0=0.1,r_1=0.115,r=\sqrt{(x-0.5)^2+(y-0.5)^2}$, and the taper function $f(r)$ is defined by $f(r)=\bigl(r_1-r\bigr)/\bigl(r_1-r_0\bigr)$. The value $\gamma=1.4$ is used. 

Panels in Figure \ref{Rotor3dContour} exhibit contour plots in $x$-$y$ plane of the density, magnetic pressure and Mach number at the final time $t=0.15$. Contour slices are taken at $z=0.5$. The problem is solved on a $128^3$ grid resolution using the Roe solver with PPM. MC limiter is used for the PPM reconstruction.
For all cases 40 equally spaced contour lines are plotted. 
All of the contour plots show that our 3D results correspond very closely to the underlying 2D solutions (e.g., see [\citenum{LeeDeane2009}]). As reported in [\citenum{Toth2000}] one important feature to observe in this problem is to check the oval contours of Mach number near the center. As illustrated, the contour lines
are symmetrical and well preserved with our choice of CFL=0.95.

\begin{figure}[htbp]
  \centerline{   
    {\subfigure[Density contour at $z=0.5$ ranging between 0.4540 and 14.82] {\epsfig{file=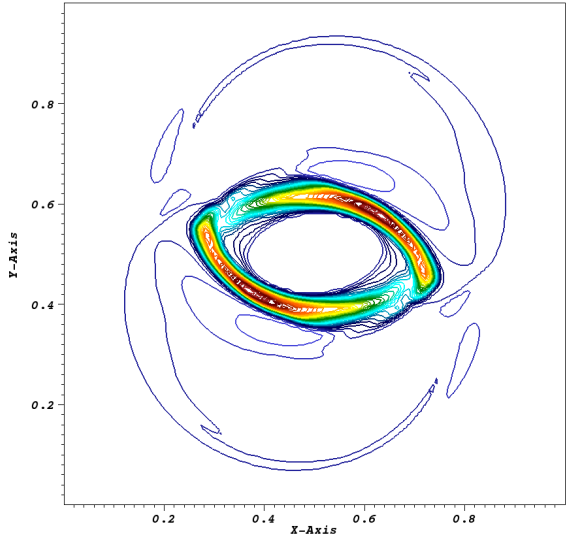, height=3.5 in, width=3.5 in}}}
    {\subfigure[Magnetic pressure contour at $z=0.5$ ranging between 0.009705 and 3.171]{\epsfig{file=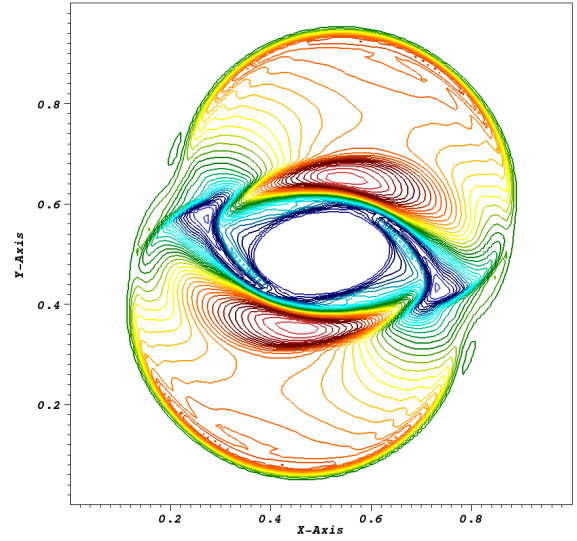, height=3.5 in, width=3.5 in}}}
    }
   \centerline{
    {\subfigure[Mach number contour at $z=0.5$ ranging between $3.268\times10^{-5}$ and 5.938] {\epsfig{file=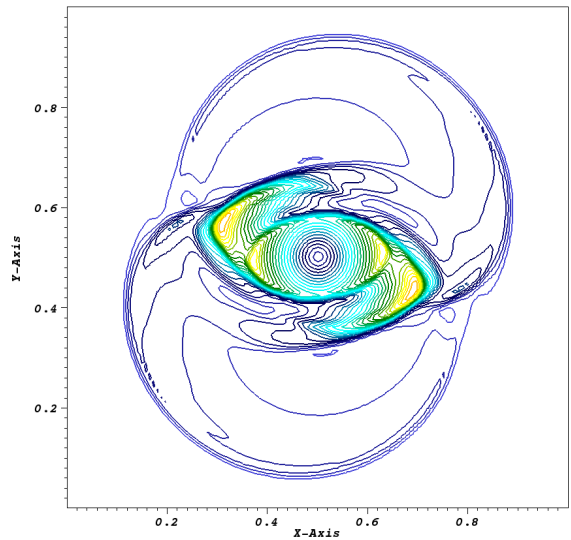, height=3.5 in, width=3.5 in}}}
    }
\caption{The rotor problem with a resolution of $128^3$ at $t=0.15$. In all cases, 40 equally spaced contour lines are plotted.}
\label{Rotor3dContour}
\end{figure}

\subsection{Cloud \& Shock Interaction}
\label{sec4:cloudShock}
In the next test problem we consider the interaction of a high density
cloud with a strong shock wave, originally studied by Dai and Woodward 
[\citenum{DaiWoodward1998}] and often referred to as the 
cloud-shock interaction problem.
This problem aims to test
the code robustness in solving
flow conditions such as high supersonic Mach numbers in the pre-shock and the post-shock regions, 
wide ranges of plasma beta values across the front/rear of the cloud, and strong shear flows
[\citenum{Toth2000,LeeDeane2009,MiniatiMartin2011,Helzel2011}].

Our computational domain is a cube, spanning from -0.5 to 0.5 in all three directions 
and is resolved on $128^3$ grid cells. 
Supersonic inflow boundary conditions are imposed at the lower boundary $x=-0.5$, 
while outflow conditions are used elsewhere.
The initial condition has different left and right states, separated by an initial discontinuity 
at $x=0.1$, given by
\begin{eqnarray}
\left(\rho,u,v,w,B_x,B_y,B_z,p\right)=
\cases{\left(3.86859,0,0,0,0,2.1826182,-2.1826182,167.345\right) 
	\;\;\;\mbox{if} \;\; x \leq 0.1,\cr
       \left(1,-11.2536,0,0,0,0.56418958,0.56418958,1    \right) 
	\;\;\;\mbox{if} \;\; x > 0.1.}
\end{eqnarray}
The high density cloud is located on the right side of the domain, and
has a spherical envelope defined by $(x-0.3)^2 + y^2 + z^2 = 0.15^2$.
A uniform density $\rho=10$ and pressure $p=1$ are fixed in the inner
region of the cloud, and $\gamma=5/3$ is used everywhere. 
The velocity and the magnetic fields are the same as the surrounding right state plasma values. 
The simulation is carried out to a final time $t=0.06$ using the WENO5 reconstruction scheme and the
Roe Riemann solver. We used the van Leer's slope limiter for limiting characteristic variables in the WENO5
reconstruction.

Figure \ref{CloudShockInteraction} shows density (plotted in the top half using a red color scheme) and 
magnetic pressure (plotted in the bottom half using a blue color scheme) at $t=0.06$. 
The main features of the cloud-shock interaction process are well captured, in that 
the temporal evolution of the high density cloud produces disrupted shapes
as the cloud moves into the plane shock on the left.

Simulations of this problem are often performed using a rather diffusive set of numerical options 
such as the minmod slope limiter, HLL-type Riemann solvers, or lower values of CFL.
For example, as noted by T\'{o}th, dimensionally-split MHD algorithm can easily fail due to unphysical states
(e.g., negative pressure or density) arising in consequence of the strong interaction between the shock and the cloud. 
By contrast, the 3D USM scheme utilizing the full 3D CTU algorithm and the upwind-MEC scheme
can run this simulation successfully without relying on such numerically dissipative choices. 
Despite our choice of the van Leer's limiter for WENO5, and of the Roe Riemann solver using CFL=0.95,
the final time step is reached successfully without giving rise to any numerical instabilities.

\begin{figure}[htbp]
  \centerline{
    {\subfigure[Density and magnetic pressure at $t=0.06$]{\epsfig{file=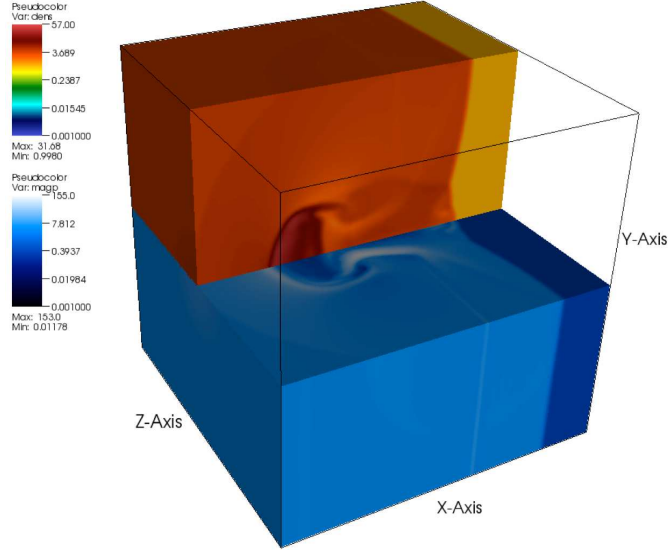, height=3.2 in, width=3.5 in}}}
    }
\caption{The 3D MHD interaction between the high density cloud and shock 
structures resolved on $128^3$ grid using the Roe Riemann solver and the 5th order WENO scheme.
Plotted are density (denoted as "dens" in the legend) in the top half and magnetic pressure (denoted as "magp") in the bottom half.}
\label{CloudShockInteraction}
\end{figure}

\subsection{MHD Blast Wave}
\label{sec4:blastWave}
The last test case is the 2D MHD spherical blast wave problem of
Zachary {\em et al.} [\citenum{ZacharyMalagoliColella1994}]. We presented
our 2D results in [\citenum{LeeDeane2009}] and extend the problem to 3D here.
We test three different configurations, differing by the initial strength of
magnetic field in $x$-direction, each leading to strong shock formation and propagation.

The computational domain is a unit cube  $[-0.5, 0.5]\times[-0.5,0.5]\times[-0.5,0.5]$
with a grid resolution of $128^3$. The ambient gas is initialized as
\begin{equation}
\rho=1, \;p  =0.1, \; \mathbf B=(B_{x_0},0,0)^T,
\end{equation}
where the three simulations have initial values of $B_{x_{0}}$ given by
$B_{x_0}=0$, $B_{x_0}=\frac{50}{\sqrt{4\pi}}$ and $B_{x_0}=\frac{100}{\sqrt{4\pi}}$.
At the center of the domain, a spherical region of radius $r=0.1$ 
is initialized with a very strong pressure $p=1000$.
The non-zero values of $B_{x_0}=\frac{50}{\sqrt{4\pi}}$ and $\frac{100}{\sqrt{4\pi}}$ produce
very low-$\beta$ ambient plasma states, $\beta=1\times10^{-3}$ and
$2.513\times10^{-4}$ respectively. Through these low-$\beta$ ambient
states, the explosion initially emits almost spherical fast magneto-sonic shocks that
propagate with the fastest wave speed. The flow has $\gamma=1.4$.

Shown in Figures \ref{BlastBx0}--\ref{BlastBx100} are (a) density (plotted in the top half) and 
magnetic pressure (plotted in the bottom half) and (b) contour plots of gas pressure (top half) and
total velocity $U=\sqrt{u^2+v^2+w^2}$ (bottom half) at time $t=0.01$. The contour slice plots are the $x$-$y$ planes taken at $z=0$.

This problem is susceptible to a type of shock wave instability known as the carbuncle phenomenon
[\citenum{Quirk1994}]. 
The carbuncle instability takes place in multidimensional numerical solutions
when using a less dissipative, 1D based (rather than the multidimensional based [\citenum{Balsara2010}])
Roe-type Riemann solver, in the regions where a planar shock is aligned to the grid.
The cause of this instability is the lack of numerical diffusivity  added to the Roe-type fluxes 
perpendicular to the grid-aligned shock, resulting in a growth of small amplitude noise in the transverse direction.
There are several approaches to fix the instability [\citenum{Sanders1998, PandolfiDAmbrosio2001, Hanawa2008,StoneGardinerEtAlAthena2008}] 
which all basically provide a similar mechanism to add extra numerical diffusion in the transverse direction.
Here we use a hybrid Riemann solver that appropriately combines Roe and HLLE depending on the strength of shocks.
In this approach, the HLLE solver is adaptively used only in strong shock fronts detected by a shock switch [\citenum{BalsaraSpicer1999}]; the Roe solver is used elsewhere. The second-order accurate MUSCL-Hancock
scheme is used for the normal predictor calculations. 
We also employ a hybrid-type of slope limiter  that combines MC limiter for linearly degenerate waves 
(i.e.,  Alfv\'{e}n and entropy waves) and the minmod limiter for genuinely nonlinear waves (i.e., magneto-sonic fast and slow waves). 
This hybrid limiter approach provides an added robustness and accuracy by using 
a compressive limiter (such as MC and van Leer's) for crisper representation of the linear waves,
whereas a diffusive limiter (such as minmod) for the self-steepening nonlinear waves [\citenum{Balsara2003}].

The case $B_{x_0}=0$ is illustrated in Figure \ref{BlastBx0}. The carbuncle phenomenon can appear to be
stronger in this hydrodynamic limit than when $B_{x_0}\ne 0$. Using the hybrid Riemann solver, however,
we do not see any artifacts at the shock fronts that are aligned to the grid axes. 
In the absence of magnetic field the explosion propagates the shock wave
spherically in all radial directions, as exhibited in the contour plots in Figure \ref{BlastBx0} (b).

\begin{figure}[htbp]
  \centerline{   
    {\subfigure[Density and magnetic pressure at $t=0.01$]
	{\epsfig{file=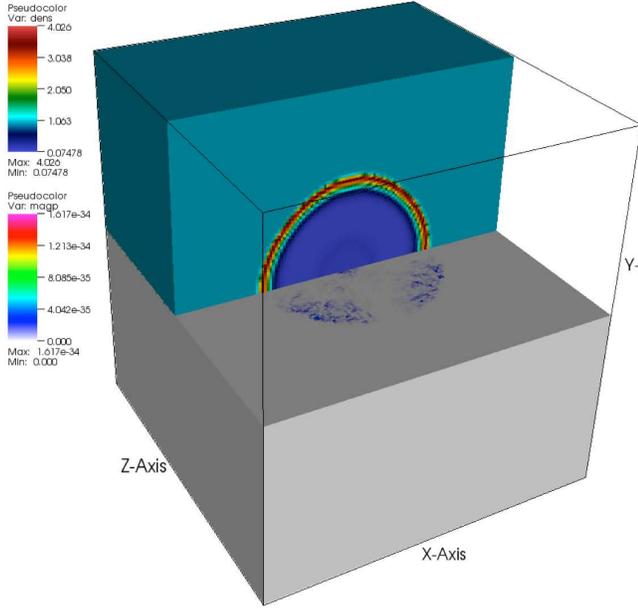 , height=3.2 in, width=3.5in}}}
   {\subfigure[Contours of gas pressure and total velocity at $t=0.01$ in the $x$-$y$ plane at $z=0$]
	{\epsfig{file=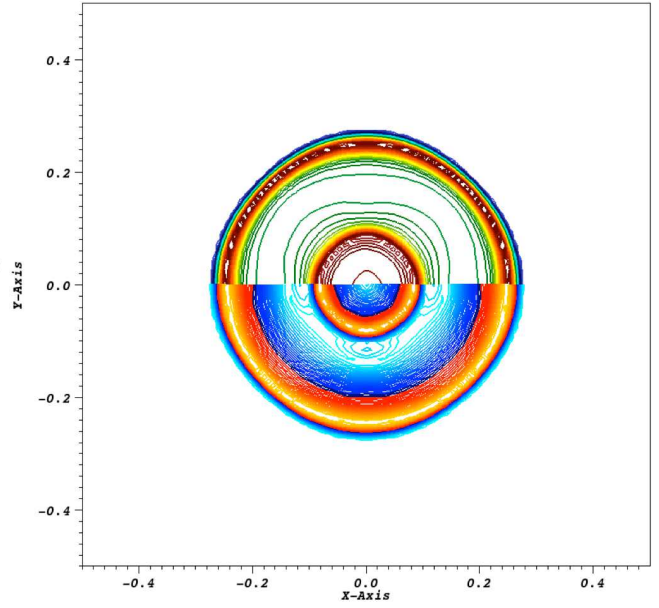, height=3.2 in, width=3.4in}}}
    }
\caption{Results of the blast problem simulation with $B_x=0$ using a hybrid Riemann solver. 
In (a), density (denoted as "dens" in the legend) is plotted at the top half. Magnetic pressure (denoted as "magp" in the legend)
is plotted at the bottom half and is represented as small values that are numerical noise. In (b), 40 contour lines are plotted for gas pressure (top half) between 0.1 and 73.62 and total velocity (bottom half) between 0 and 8.810.}
\label{BlastBx0}
\end{figure}

In Figure {\ref{BlastBx50}} the intermediate magnetic field strength case with $B_x=\frac{50}{\sqrt{4\pi}}$ is
shown. The explosion becomes anisotropic because of the 
non-zero magnetic field strength in $x$-direction. 
The intermediate value of $B_x$ still permits shock wave propagation in the $y$-direction,
so that the overall spherical shape is not radically distorted. Nonetheless, the
development of the elongated wave structures in the direction parallel to
the $B_x$ field is evident compared to the hydrodynamic limit case in Figure \ref{BlastBx0}.

 \begin{figure}[htbp]
  \centerline{   
    {\subfigure[Density and magnetic pressure at $t=0.01$]
	{\epsfig{file=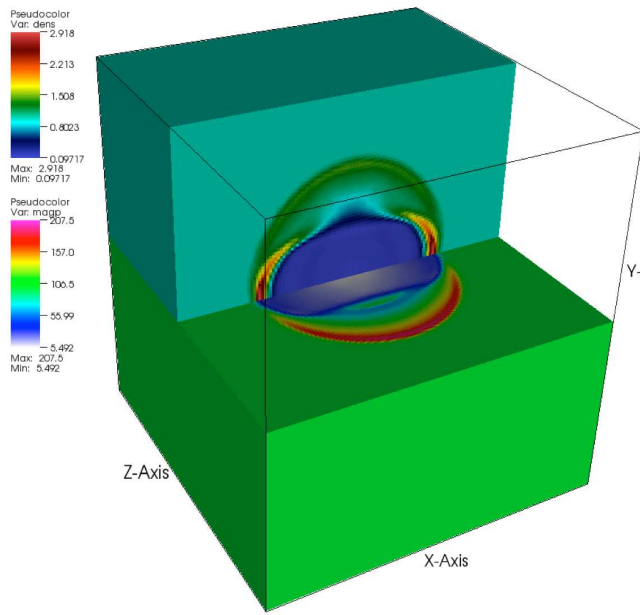 , height=3.2 in, width=3.5in}}}
   {\subfigure[Contours of gas pressure and total velocity at $t=0.01$ in the $x$-$y$ plane at $z=0$]
	{\epsfig{file=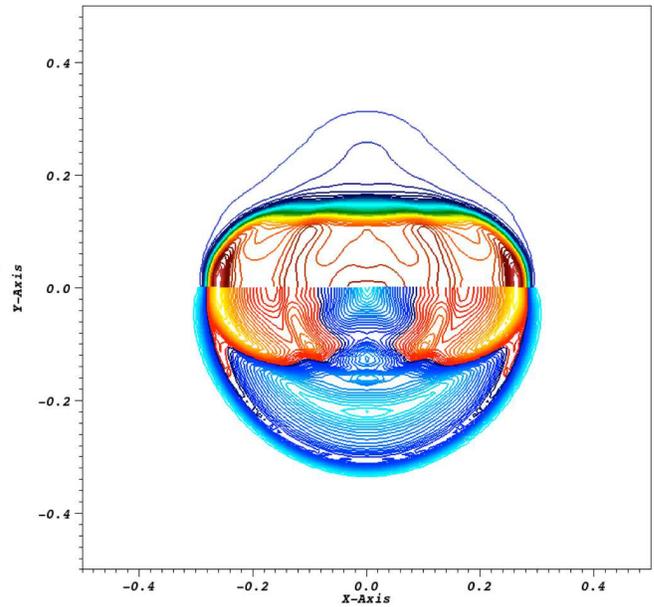, height=3.2 in, width=3.4in}}}
    }
\caption{Results of the blast problem simulation with $B_x=\frac{50}{\sqrt{4\pi}}$ using a hybrid Riemann solver. In (a), density (denoted as "dens" in the legend) is plotted at the top half. Magnetic pressure (denoted as "magp" in the legend) is plotted at the bottom half. In (b), 40 contour lines are plotted for gas pressure (top half) between 0.009981 and 106.6 and total velocity (bottom half) between 0 and 11.84.}
\label{BlastBx50}
\end{figure}

Finally, Figure {\ref{BlastBx100}} illustrates the strongest magnetic field case, $B_x=\frac{100}{\sqrt{4\pi}}$. 
The explosion now becomes highly anisotropic.
This strong anisotropic behavior is well shown in Figure {\ref{BlastBx100}}(b) 
in that the displacement of gas in the transverse
$y$-direction is increasingly inhibited and hydrodynamical shocks propagate
almost entirely in the $x$-direction parallel to $B_x$. It is also
evident that several weak magneto-sonic waves are radiated
transverse to the $x$-direction. This process continues until total pressure
equilibrium is reached in the central region.

\begin{figure}[htbp]
  \centerline{   
    {\subfigure[Density and magnetic pressure at $t=0.01$]
	{\epsfig{file=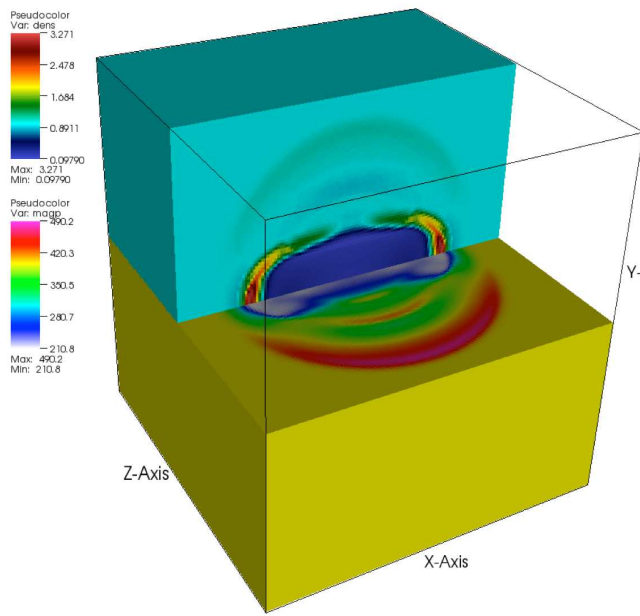 , height=3.2 in, width=3.5in}}}
   {\subfigure[Contours of gas pressure and total velocity at $t=0.01$ in the $x$-$y$ plane at $z=0$]
	{\epsfig{file=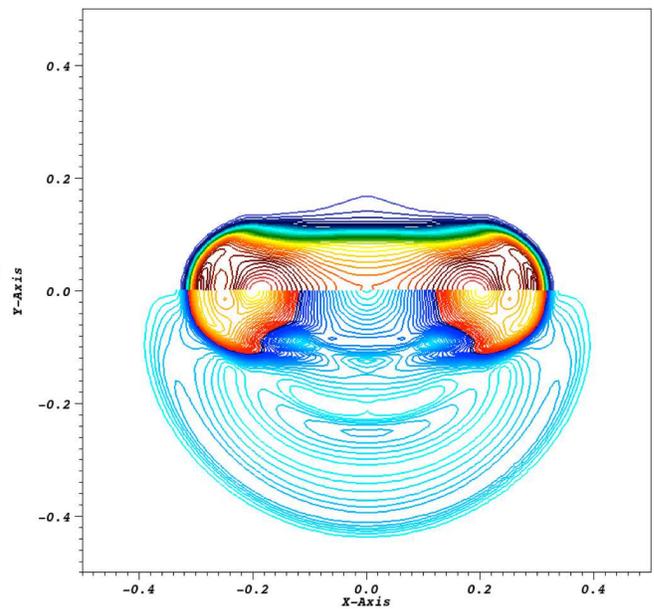, height=3.2 in, width=3.4in}}}
    }
\caption{Results of the blast problem simulation with $B_x=\frac{100}{\sqrt{4\pi}}$ using a hybrid Riemann solver. In (a), density (denoted as "dens" in the legend) is plotted at the top half. Magnetic pressure (denoted as "magp" in the legend) is plotted at the bottom half. In (b), 40 contour lines are plotted for gas pressure (top half) between 0.009161 and 202.9 and total velocity (bottom half) between 0 and 15.97.}
\label{BlastBx100}
\end{figure}

Balsara [\citenum{Balsara2003}] pointed out that maintaining positivity of pressure 
becomes challenging due to the strong wave propagation
oblique to the mesh. Such unphysical pressures
can distort contours, especially near the outer
boundary where a large and unphysical drop in pressure takes place
immediately ahead of the shock. In our calculations, the pressure remains
always positive throughout the simulations, evidence that our 3D MHD scheme is
very robust and accurate with our choice of high CFL=0.95.

\section{Conclusion}
\label{sec:conclusion}
We summarize several key features described in this paper.
First, the 3D USM scheme has been introduced, developed and studied. 
The method is a 3D extension of the 2D USM algorithm 
[\citenum{LeeDeane2009}] which employs characteristic analysis to account for contributions
of both normal and transverse MHD fluxes in a truly unsplit fashion.
Therefore they do not need intermediate Riemann solves to correct
the normal predictor states with the transverse flux updates as in the usual
12-solve CTU algorithm [\citenum{Saltzman1994,MiniatiMartin2011}].
Our approach of using characteristic analysis provides 
computational efficiency by storing the eigensystem
evaluations when computing each normal direction and re-using them in the transverse
flux calculations.

We introduced two different methods, the reduced and full 3D CTU schemes. 
The reduced scheme
can be considered as a straightforward extension of the 2D CTU algorithm 
of Colella [\citenum{Colella1990}], and is analogous to the
6-solve algorithm in 3D by Gardiner and Stone [\citenum{GardinerStone2008}].
Although the reduced CTU scheme has a simple implementation
for 3D, its
stability limit is bounded by a CFL number less than 0.5. 
The full CTU scheme significantly improves
this limited stability range and can utilize the maximum stability range of 
CFL number close to 1. This was achieved by taking into account
the second- and third-order cross derivative terms in computing 
intermediate states at $n+\frac{1}{3}$ and $n+\frac{1}{2}$. Our full CTU scheme
thus includes the multidimensional upwind information that is crucial
to provide the full CFL limit. We also showed that the relative
CPU cost of the full scheme compared to the reduced scheme is
less than 1, indicating the cost efficiency of the full CTU scheme.
The multidimensional MHD terms are 
also properly included in 
both normal and transverse directions.

Second, we extensively investigated the lack of numerical dissipation mechanisms
in the existing CT algorithms, especially when there is a biased direction in
advecting magnetic fields. In the small angle advection tests in 2D and 3D,
we showed that the field loop simply can fail to be cleanly advected, and become distorted into
non-circular or non-cylindrical shapes in most CT schemes.
By contrast, the upwind-MEC scheme,
by incorporating upwind information
adds the needed numerical dissipation when taking the arithmetic average
in CT. The algorithm enhances the previous MEC scheme [\citenum{LeeDeane2009}]
in that upwind-MEC maintains consistency of plane-parallel and
grid-aligned flows [\citenum{GardinerStone2005}].

The results of the test problems in Section \ref{sec:results} give considerable confidence in our
scheme for use as a robust and reliable second-order, finite-volume 
3D MHD algorithm. The methods developed in this paper for the 
3D USM scheme preserve the divergence-free constraint  
without any evidence of numerical instability or accumulation of
unphysical errors using a very high CFL number close to 1.
The suite of test problems presented in
this study include several stringent setups 
can be particularly challenging for
MHD algorithms. The scheme has been thoroughly tested and has been shown to
perform very well, providing confidence in its ability to correctly simulate a wide
range of MHD phenomena.

The 3D USM scheme presented here has been implemented on 
both uniform and AMR grids. It has been integrated and tested in the
official FLASH4 release of the Flash Center for Computational Science 
at the University of Chicago [\citenum{Flash}].

\section{Acknowledgments}
This work was supported in part at the University of
Chicago  by the US Department of Energy (DOE) under contract B523820
to the NNSA  ASC/Alliances Center for Astrophysical Thermonuclear
Flashes; the Office  of Advanced Scientific Computing Research, Office
of Science, US DOE,under contract DE-AC02-06CH11357; the US DOE NNSA
ASC through the Argonne Institute for Computing in Science under field
work proposal 57789; and the US National Science Foundation under
grant PHY-0903997. 

The software used in this work was developed in part by the 
DOE NNSA ASC- and DOE Office of Science ASCR-supported Flash Center for 
Computational Science at the University of Chicago.

The author gratefully acknowledge the FLASH group for help and for supporting
the current work. The author also thank D. S. Balsara and 
anonymous referees for very helpful suggestions and comments on the manuscript.

\bibliography{mhdPaper3d}
\bibliographystyle{plain}
\end{document}